\documentclass[11pt,onecolumn]{IEEEtran}
\pdfoutput=1
\usepackage{cite}

\usepackage{amssymb}
\usepackage{amsmath}
\usepackage{times}
\usepackage{theorem}
\usepackage{graphicx}
\graphicspath{{./graphics/}{.}}
\usepackage{color}

\long\def\symbolfootnote[#1]#2{\begingroup
\def\thefootnote{\fnsymbol{footnote}}\footnote[#1]{#2}\endgroup}

\newcommand{\Exp}[1]{\mathbb{E}\left\{#1\right\}}
\newcommand{\Expected}{\mathbb{E}}

\newtheorem{thm}{Theorem}

\newcommand{\bsm}[1]{{\boldsymbol #1}}
\newcommand{\hatbsm}[1]{\hat{\boldsymbol #1}}
\newcommand{\barbsm}[1]{\bar{\boldsymbol#1}}
\newcommand{\hbarbsm}[1]{\hat{\bar{\boldsymbol #1}}}
\newcommand{\norm}[1]{\Vert#1\Vert}

\begin{document}

\title {Widely Linear vs. Conventional Subspace-Based Estimation of SIMO Flat-Fading Channels: Mean-Squared Error Analysis}

\author{Saeed Abdallah and Ioannis N. Psaromiligkos*\thanks{The authors S. Abdallah and I. Psaromiligkos are with McGill University, Department of Electrical and Computer Engineering, 3480 University Street, Montreal, Quebec, H3A 2A7, Canada, Email: saeed.abdallah@mail.mcgill.ca; yannis@ece.mcgill.ca, phone: +1 (514) 398-2465, fax: +1 (514) 398-4470.}\thanks{*Corresponding author.}}

\maketitle

\begin{abstract}

We analyze the mean-squared error (MSE) performance of widely linear (WL) and conventional subspace-based channel estimation for single-input multiple-output (SIMO) flat-fading channels employing binary phase-shift-keying (BPSK) modulation when the covariance matrix is estimated using a finite number of samples. The conventional estimator suffers from a phase ambiguity that reduces to a sign ambiguity for the WL estimator. We derive closed-form expressions for the MSE of the two estimators under four different ambiguity resolution scenarios. The first scenario is optimal resolution, which minimizes the Euclidean distance between the channel estimate and the actual channel. The second scenario assumes that a randomly chosen coefficient of the actual channel is known and the third assumes that the one with the largest magnitude is known.  The fourth scenario is the more realistic case where pilot symbols are used to resolve the ambiguities. Our work demonstrates that there is a strong relationship between the accuracy of ambiguity resolution and the relative performance of WL and conventional subspace-based estimators, and shows that the less information available about the actual channel for ambiguity resolution, or the lower the accuracy of this information, the higher the performance gap in favor of the WL estimator.

\end{abstract}

\section{Introduction}

Subspace-based estimation is one of the most popular approaches for blind channel estimation. Originally proposed in~\cite{moulines95}, subspace-based methods estimate the unknown channel by exploiting the orthogonality between the signal and noise subspaces in the covariance matrix of the received signal, offering a convenient tradeoff between performance and computational complexity~\cite{advances_subspace}. They potentially outperform methods that are based on higher-order statistics when the number of available received signal samples is limited, since second-order statistics (SOS) can be estimated more robustly in such conditions~\cite{prof_champagne}.
Subspace-based channel estimation has been applied in a wide variety of communication systems, including single-carrier systems~\cite{moulines95}, multicarrier (OFDM) systems~\cite{muquet02,roy03}, and spread spectrum (CDMA) systems~\cite{bensley96,liu96,xu04}. 

It has become a well-known fact that widely linear (WL) processing~\cite{picinbono95}, which operates on both the received signal and its complex conjugate, can improve the performance of SOS-based algorithms when the signal under consideration is improper. Improper signals, which are characterized by a non-zero pseudo-covariance, result when communication systems employ real modulation schemes such as amplitude-shift-keying (ASK), binary phase-shift-keying (BPSK), minimum-shift-keying (MSK) and Gaussian minimum-shift-keying (GMSK).  By augmenting the observation space, WL processing is able to access and exploit the information in the pseudo-covariance of the signal. This has prompted researchers to propose WL versions of subspace-based channel estimation algorithms in order to improve channel estimation accuracy in communication systems that employ real signaling. In the context of DS-CDMA systems, a WL subspace-based channel estimation algorithm was proposed in~\cite{zarifi06} for the case of BPSK modulation. In addition to exhibiting superior mean-squared error (MSE) performance, the WL algorithm was able accommodate almost twice as many users. Similar observations were made in the context of multicarrier CDMA in~\cite{gelli6blind}. A WL subspace-based algorithm was developed for the estimation of single-input single-output (SISO) FIR channels with improper input signals in~\cite{siso_subspace}, showing superior mean-squared error (MSE) performance to the conventional subspace-based method implemented using an oversampling factor of 2. WL subspace-based channel estimation was also used in the context of interference-contaminated OFDM systems in~\cite{multicarrier05}.

Conventional subspace-based channel estimation suffers from an inherent phase ambiguity which can only be resolved through the use of additional side information. An important advantage of WL estimation is that it reduces the phase ambiguity into a sign ambiguity~\cite{gao07, multicarrier05}, which is intuitively easier to resolve. For a meaningful comparison of WL and conventional channel estimation, both the phase ambiguity and the sign ambiguity have to be resolved. Unfortunately, there is no agreement in the literature on how to resolve the two ambiguities. In many works on blind channel estimation, it is assumed that the one of the channel coefficients, typically the first one, is known~\cite{roy02,roy03,qiu96}. This assumption was used to resolve the ambiguities of the conventional and WL estimator in~\cite{multicarrier05}. In~\cite{siso_subspace}, the channel coefficient with the largest magnitude was assumed known. Other works on WL subspace-based estimation~\cite{zarifi06, gao07} do not explicitly mention how they resolve the phase and sign ambiguities.

In this work, we consider the problem of blind estimation of single-input multiple-output (SIMO)
flat-fading channels~\cite{gorokhov03, Lapidoth06}. We assume that BPSK modulation is used, resulting in an improper
received signal. Our goal is to examine whether and under what conditions WL subspace-channel estimation constitutes an appealing alternative to conventional approaches. We pay special attention to the practical situation where estimation is performed with a finite number of received samples, and we derive highly accurate closed-form expressions for the MSE performance of the conventional and WL channel estimation algorithms. Our MSE analysis explicitly takes into account the effects of phase and sign ambiguity resolution by considering four different approaches to resolving the said ambiguities. The first approach we consider is optimal phase and sign ambiguity resolution where the applied phase and sign correction minimizes the Euclidean distance between the channel estimate and the actual channel. For this case, we also derive a closed-form expression for the probability that the WL estimator outperforms the conventional one, when the statistics of the channel are taken into account. The second approach assumes that a randomly chosen channel coefficient is known, and the third approach assumes that the channel coefficient
with the largest magnitude is known. For the latter, we also derive a lower bound on the probability
that the WL estimator outperforms the conventional one. Interestingly, the relative performance of the
two estimators is different depending on which of the above three approaches is adopted. Under optimal correction, the conventional
estimator slightly outperforms the WL estimator. However, the WL estimator is significantly better than
the conventional estimator when the  first channel coefficient is assumed known, and slightly better when the channel coefficient with the largest magnitude is assumed known. All these approaches assume that certain information about the actual channel is perfectly known at the receiver. In
practice, however, pilot symbols have to be used to estimate such information, and the incurred estimation
error will contribute further to the resulting MSE. We therefore consider a fourth approach in which a limited number of pilot symbols is used to resolve the ambiguities of the two estimators, and we derive the
corresponding MSE expressions. As it turns out, the WL estimator is significantly better, coming very
close to optimal performance even when a single pilot symbol is used. This conforms with the intuition that sign
ambiguity is much easier to correct than phase ambiguity. The four scenarios demonstrate that the less
information available about the actual channel for ambiguity resolution, or the less accurate this information is, the higher the
performance gap becomes in favor of the WL estimator.

Our work is the first to present a thorough analytical study on the relative MSE performance of conventional and WL subspace-based estimators. To the best of our knowledge, we are the first to observe and analyze the close relationship between the accuracy of phase and sign ambiguity resolution and the relative performance of WL and conventional subspace-based estimators and to analyze theoretically the practical scenario when pilots are used to resolve the phase and sign ambiguities. As such, our work offers new and unique insights into the relative performance of conventional and WL subspace-based channel estimation algorithms.

The remainder of the paper is organized as follows. In Section~\ref{system_model}, we introduce our system model and the conventional subspace-based channel estimation algorithm for the SIMO flat-fading channel model. The WL subspace-based channel estimation algorithm is developed in Section~\ref{WL_estimator}. The MSE performance of the conventional estimator under finite sample size is derived in Section~\ref{section_mse_conventional} for the four scenarios of phase ambiguity resolution. We derive the MSE performance of the WL estimator for the corresponding four sign ambiguity resolution scenarios in Section~\ref{MSE_performance_WL}. In Section~\ref{relative_performance}, we derive the probability that the WL estimator outperforms the conventional under optimal ambiguity resolution, as well as a lower bound on the probability that the WL estimator outperforms the conventional one when the channel coefficient with the largest magnitude is known. In Section~\ref{simulation_results}, we use Monte-Carlo simulations to verify the accuracy of our analytical results and to compare the performance of the two estimators in the four different cases. Finally, we present our conclusions in Section~\ref{conclusions}.

\section{System Model and Background}
\label{system_model}
\begin{figure}[t]
\centering
\includegraphics[width=3.5in]{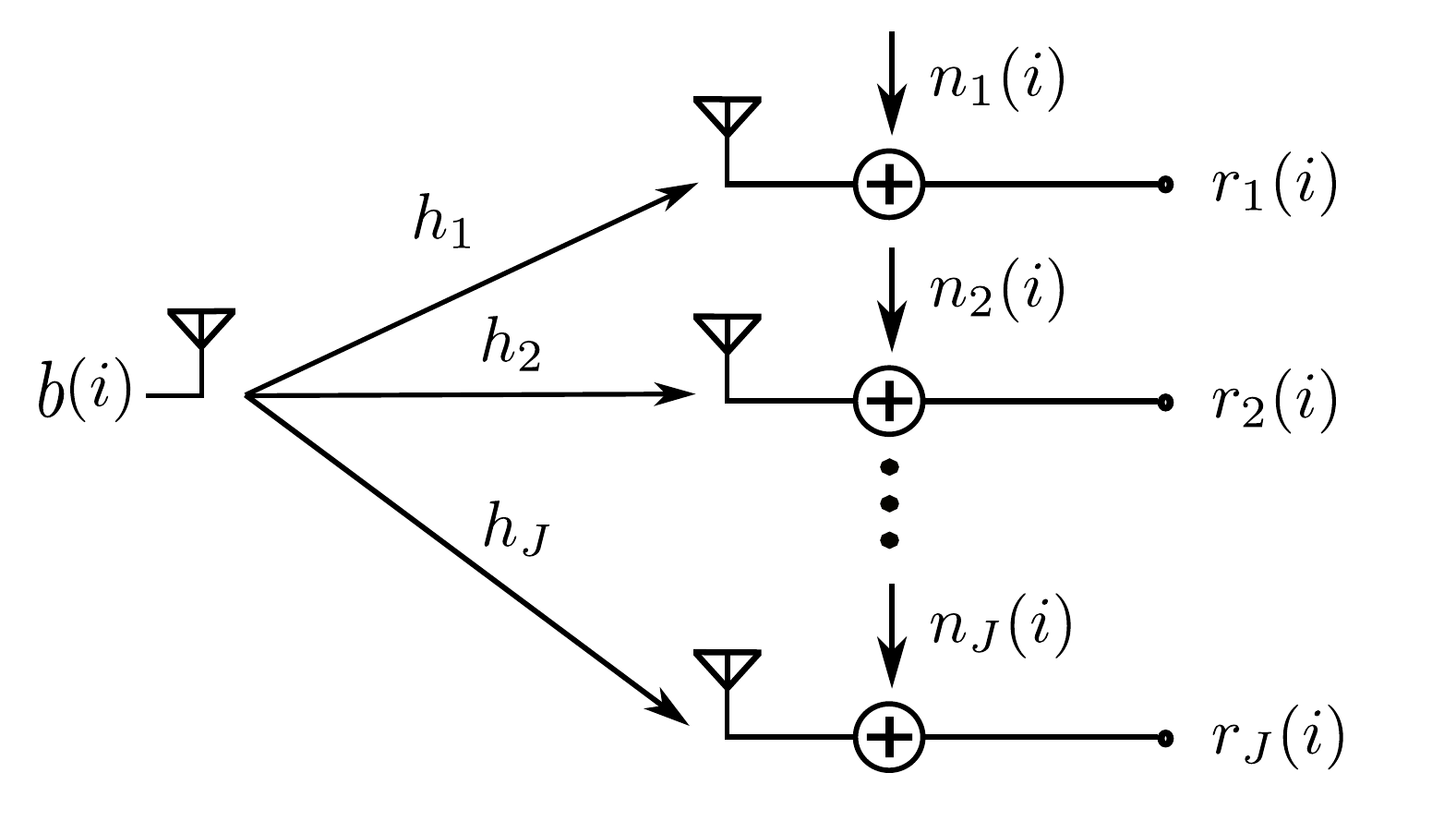}
\caption{The single user SIMO Communications Model.}
\label{fig_transmission_scheme1}
\end{figure}

We consider the single user SIMO flat-fading model used in~\cite{Lapidoth06,dogandzic04,pham08} and illustrated in Fig.~\ref{fig_transmission_scheme1}. In the \textit{i}th symbol period, the transmitter sends a binary-phase-shift-keying (BPSK) data symbol $b(i)$ taking the values $\pm 1$ with equal probability. For a $J$-antenna receiver, the corresponding $J\times1$ received vector is
\begin{equation}
\label{sm:received_vector}
\bsm{ r}(i)=b(i){\bsm{g}}+\bsm{ n}(i)=b(i)\norm{\bsm{g}}\bsm{h}+\bsm{ n}(i),
\end{equation} 
where ${\bsm{g}}\triangleq[g_1,\hdots,g_J]^{T}$ is the unknown vector of fading channel coefficients, $\bsm{ h}\triangleq{\bsm{g}}/\norm{\bsm{g}}=[h_1,\hdots,h_J]$ is the normalized (to unit-norm) version of the channel, and $\bsm{ n}(i)$ is the complex additive white Gaussian noise (AWGN) vector with mean zero and covariance $\Expected[\bsm{ n}(i)\bsm{ n}(i)^H]=\sigma^2\bsm{ I}$. The fading coefficients $g_1,\hdots,g_J$ are modelled as independent and identically distributed (i.i.d.) $\mathcal{CN}(0,\gamma^2)$\footnote{The notation $\mathcal{CN}(\mu,\gamma^2)$ is used to refer to the complex Gaussian distribution with mean $\mu$ and variance $\gamma^2$.} and are assumed to remain fixed throughout the estimation period. The transmit SNR in dB is $-10\log(\sigma^2)$. It is well known that SOS-based algorithms are blind to the channel magnitude $\norm{\bsm{g}}$, i.e., they can only estimate the direction and not the norm of $\bsm{g}$. For this reason, we will focus on estimating $\bsm{ h}$, and, with a slight abuse of terminology, we will also refer to the elements of vector $\bsm{ h}$ as fading coefficients.

Let $\bsm{ R}\triangleq\Expected\{\bsm{ r}(i)\bsm{ r}(i)^{H}\}$ be the covariance matrix of the received signal, then
\begin{equation}
\label{covariance_matrix}
\begin{split}
\bsm{ R}&={\bsm{g}}{\bsm{g}}^{H}+\sigma^2\bsm{ I}=\norm{\bsm{g}}^2\bsm{ h}\bsm{ h}^{H}+\sigma^2\bsm{ I}.
\end{split}
\end{equation}
As we can see from~\eqref{covariance_matrix}, the matrix $\bsm{ R}$ has two distinct eigenvalues, $\lambda_{1}=(\norm{\bsm{g}}^2+\sigma^{2})$ and $\lambda_{2}=\sigma^{2}$, the latter having a multiplicity of $J-1$. The normalized channel $\bsm{ h}$ is an eigenvector of $\bsm{ R}$ corresponding to the largest eigenvalue. If we let $\bsm{ u}$ be a unit-norm eigenvector of $\bsm{ R}$ corresponding to $\lambda_{1}$, then $\bsm{ u}$ has the form $\bsm{ u}=\bsm{ h}e^{\jmath\phi}$, where $\jmath=\sqrt{-1}$ and $\phi\in[0,2\pi)$ is an arbitrary angle which represents the ambiguity in the phase of the channel estimate which is common to all SOS-based estimators. 
In practice, $\bsm{ R}$ is not known beforehand and is commonly replaced by its sample-average estimate $\hatbsm{ R}=\frac{1}{N}\sum_{i=1}^{N}\bsm{ r}(i)\bsm{ r}(i)^{H}$, using a finite sample of $N$ received vectors, $\bsm{r}(1),\hdots,\bsm{r}(N)$. In this case, the channel is estimated by the unit-norm eigenvector corresponding to the largest eigenvalue of $\hatbsm{ R}$. We denote by $\hat{\bsm{ u}}\triangleq[\hat{u}_1,\hdots\hat{u}_J]^{T}$ the subspace-based channel estimate prior to phase correction. For a meaningful study of the performance of the subspace-based estimator, it is essential to resolve the phase ambiguity by applying an appropriate phase shift to $\hat{\bsm{ u}}$. The most common convention is to assume that one of the channel coefficients, typically the first one, is known~\cite{roy02,roy03,qiu96}, and rotate the vector $\hat{{\boldsymbol u}}$ such that the corresponding component has the same phase. This approach may lead to inaccurate results when the chosen coefficient happens to have a very small magnitude~\cite{wang03}. An alternative approach which leads to more accurate ambiguity resolution is to assume that the known channel coefficient is the one with the largest magnitude~\cite{siso_subspace}. In both cases, however, the phase shift is suboptimal because it uses information from only one channel component. In fact, the optimal phase shift which minimizes the Euclidean distance between the channel estimate and the actual channel is the phase of their inner product~\cite{wang03}. 
In practice, however, none of the information used in these three approaches is readily available at the receiver, and pilot symbols have to be used to resolve the phase ambiguity. Obviously, the choice of how to resolve the phase ambiguity will affect the observed mean-squared error (MSE) performance. When training pilots are used, the error in estimating the desired phase shift will further contribute to the MSE.

In section~\ref{section_mse_conventional}, we will derive accurate closed-from expressions for the MSE error of the conventional subspace-based estimator under each of the four scenarios. 

\section{The Widely Linear Subspace-based Estimator}
\label{WL_estimator}
Due to the use of real (BPSK) modulation, the received signal is improper, and widely linear processing may be applied to access the information in the pseudo-covariance of the signal. In this section, we will develop the widely linear version of the subspace-based channel estimator described in Section~\ref{system_model}. Widely linear processing operates on an augmented vector $\tilde{\boldsymbol r}(i)$ formed by stacking the received vector ${\boldsymbol r}(i)$ and its complex conjugate ${\boldsymbol r}(i)^{*}$. The vector $\tilde{\boldsymbol r}(i)$ is given by
\begin{equation}
\tilde{\boldsymbol r}(i)\triangleq \begin{bmatrix}\ \ {\boldsymbol r}(i)\\ \ {\boldsymbol r}(i)^{*}\end{bmatrix}=b(i)
\norm{\bsm{g}}\begin{bmatrix}
\ {\boldsymbol h}\\
\ {\boldsymbol h}^{*}
\end{bmatrix}+\begin{bmatrix}
\ {\boldsymbol n}(i)\\
\ {\boldsymbol n}(i)^{*}
\end{bmatrix}=b(i)\norm{\bsm{g}}\tilde{\boldsymbol h}+\tilde{\boldsymbol n}(i),
\end{equation}
where $\tilde{\boldsymbol h}=\begin{bmatrix}
\ {\boldsymbol h}^{T}&
{\boldsymbol h}^H
\end{bmatrix}^{T}$ and $\tilde{\boldsymbol n}(i)=\begin{bmatrix}
\ {\boldsymbol n}(i)^T&{\boldsymbol n}(i)^{H}
\end{bmatrix}^{T}$.
The covariance matrix $\tilde{\boldsymbol R}$ of $\tilde{\boldsymbol r}(i)$ is given by 
\begin{equation}
\label{eq_correlation_wl}
\tilde{\boldsymbol R}\triangleq\Expected [\tilde{\boldsymbol r}(i)\tilde{\boldsymbol r}(i)^{H}]=\begin{bmatrix}\ {\boldsymbol R}&{\boldsymbol C}\\
\ {\boldsymbol C}^*&{\boldsymbol R}^*\end{bmatrix},
\end{equation}
where ${\boldsymbol C}\triangleq\Expected\{{\boldsymbol r}(i){\boldsymbol r}(i)^{T}\}$ is the pseudo-covariance of ${\boldsymbol r}(i)$. It is easy to see that the augmented channel vector $\tilde{\boldsymbol h}$ is an eigenvector of $\tilde{\boldsymbol R}$ corresponding to the largest eigenvalue, $\lambda_{w}=2\norm{\bsm{g}}^2+\sigma^2$. 

WL channel estimation reduces the inherent phase ambiguity into a sign ambiguity. This fact becomes clear by reformulating the WL subspace-based estimator into the following equivalent real representation. Let $\barbsm{ r}(i)\triangleq[\Re\{{\boldsymbol r}(i)\}^{T},\Im\{{\boldsymbol r}(i)\}^{T}]^{T}$ be the real representation of the received vector ${\boldsymbol r}(i)$, and ${\bar {\boldsymbol h}}\triangleq[\Re\{{\boldsymbol h}\}^{T},\Im\{{\boldsymbol h}\}^{T}]^{T}=[\bar{h}_1,\hdots,\bar{h}_{2J}]^{T}$ be the real representation of the channel ${\boldsymbol h}$. Finally, let 
\begin{equation}
{\boldsymbol\Psi}\triangleq\begin{bmatrix}\ {\boldsymbol I}_{J}&\phantom{x}\jmath{\boldsymbol I}_{J}\\
\ {\boldsymbol I}_{J}&-\jmath{\boldsymbol I}_{J}\end{bmatrix}.
\end{equation}
Then, $\tilde{\boldsymbol h}={\boldsymbol\Psi}\barbsm{ h}$ and $\tilde{\boldsymbol r}(i)={\boldsymbol\Psi}\barbsm{ r}(i)$, or, equivalently, $\barbsm{ h}=\frac{1}{2}{\boldsymbol\Psi}^{H}\tilde{\boldsymbol h}$ and $\barbsm{ r}(i)=\frac{1}{2}{\boldsymbol\Psi}^{H}\tilde{\boldsymbol r}(i)$. The covariance matrix of $\barbsm{ r}(i)$ is  $\barbsm{ R}\triangleq\Expected\{\barbsm{ r}(i)\barbsm{ r}(i)^{T}\}=\norm{\bsm{g}}^2\barbsm{ h}\barbsm{ h}^{T}+\frac{\sigma^2}{2}{\boldsymbol I}$, and it is related to $\tilde{\bsm{R}}$ by $\barbsm{R}=\frac{1}{4}{\boldsymbol\Psi}^{H}\tilde{\boldsymbol R}{\boldsymbol\Psi}$. Moreover, there is a 1-1 correspondence between the eigenvalues and eigenvectors of $\tilde{\boldsymbol R}$ and those of $\barbsm{ R}$. Hence, WL subspace-based channel estimation can be performed in the real domain through the eigendecomposition of $\barbsm{ R}$. We are left with a sign ambiguity because both $\barbsm{ h}$ and $-\barbsm{ h}$ are unit-norm eigenvectors of $\barbsm{ R}$ corresponding to its largest eigenvalue. For the remainder of the paper, we will use the real representation of the WL subspace-based channel estimator. 

Since $\barbsm{ R}$ is not known to the receiver beforehand, it is approximated by its sample-average estimate ${\hat{\barbsm{ R}}}=\frac{1}{N}\sum_{i=1}^{N}\bar{\boldsymbol r}(i)\bar{\boldsymbol r}(i)^{T}$. To estimate the channel, we obtain the unit-norm eigenvector $\hat{\bar{\boldsymbol u}}$ corresponding to the largest eigenvalue of $\hat{\barbsm{ R}}$. This channel estimate suffers from a sign ambiguity which requires extra information to be resolved. In~\cite{siso_subspace}, the component of largest magnitude in the real representation of the channel was assumed to be known and used to resolve the ambiguity, while in~\cite{multicarrier05}, the real representation of the channel was converted back into a complex vector and the phase of the first coefficient of the complex channel was assumed known and a phase shift was applied to match the phase of the two vectors. The approach we follow in our work is to perform sign-ambiguity correction directly to the real channel estimate. We maintain fairness in the correction of the two ambiguities (phase in the case of conventional and sign in the case of WL) by assuming that the same information is available to both estimators. Hence, we will we perform sign ambiguity resolution under the same four scenarios discussed in Section~\ref{system_model}. Closed-form expressions for the MSE in all four cases for a finite number of samples will be derived in Section~\ref{MSE_performance_WL}.

\section{MSE Performance of the Conventional Estimator}
\label{section_mse_conventional}

In this section, we derive closed-form expressions for the mean-squared error performance of the conventional subspace-based estimation under the four assumptions on phase ambiguity resolution that were briefly discussed in Section~\ref{system_model}. It is more convenient to start our analysis with the case of optimal phase correction, as the resulting MSE expression will be used in the derivation of MSE expressions for the other three cases.

\subsection{Optimal Phase Correction}
\label{optimal_phase_correction}

Let $\theta_o$ be the optimal phase shift that minimizes the Euclidean distance between the channel estimate and the true channel. It is straightforward to check that $\theta_o$ is given by
\begin{equation}
\theta_o=\arg\!\min_{\theta\in(0,2\pi]}\norm{e^{\jmath\theta}\hat{{\boldsymbol u}}-{\boldsymbol h}}^2=\angle\left(\hat{{\boldsymbol u}}^{H}{\boldsymbol h}\right).
\end{equation}
Let $\hatbsm{h}_o\triangleq e^{\jmath\theta_o}\hatbsm{u}=[\hat{h}_{o,1},\hdots,\hat{h}_{o,J}]$ be the optimally phase-corrected estimate of the channel, and let $\delta\bsm{h}\triangleq\hatbsm{h}_o-\bsm{h}$ be the corresponding error in the estimation of $\bsm{h}$ under optimal phase correction. We can decompose $\delta\bsm{h}_o$ as
\begin{equation}
\delta\bsm{h}_o\triangleq \bsm{q}_o+\alpha_o{\bsm h}, 
\end{equation}
where $\bsm{q}_o=[q_{o,1},\hdots, q_{o,J}]^{T}$ is the component of $\delta\bsm{h}_o$ that is orthogonal to $\bsm{h}$ and $\alpha_o$ is a complex scalar. Hence, 
\begin{equation}
\norm{\delta\bsm h_o}^2=\norm{\bsm q_o}^2+|\alpha_o|^2.
\end{equation}
A closed-form expression for $\Exp{\norm{\delta\bsm {h}_o^2}}$ is given by the following theorem. The proof can be found in Appendix~\ref{proof_thm_optimal}.
\begin{thm}
\label{thm_optimal}
Under optimal phase correction, $|\alpha_0|^2\ll\norm{\bsm{q}_o}^2$ and
\begin{equation*}
\label{optimal_conventional_MSE}
\Exp{\norm{\delta\bsm {h}_o}^2}\simeq\Exp{\norm{\bsm{q}_o}^2}\simeq\frac{1}{N\norm{\bsm {g}^4}}(\sigma^2\norm{\bsm{g}}^2+\sigma^4)(J-1).
\end{equation*}
\end{thm}
As we shall see shortly, the MSE under optimal phase correction lower bounds the MSE for the other three scenarios. Since the term $\alpha_o\bsm{h}$ has a negligible impact on the estimation error $\delta\bsm {h}_o$, we will assume for the remainder of our work that $\alpha_o=0$.

\subsection{Suboptimal Phase Correction}
\label{suboptimal_phase_correction}
The phase ambiguity can be resolved suboptimally if only one of the channel coefficients is known. Let $h_{\ell},\ \ell\in\{1,\hdots,J\}$, be the known channel coefficient. The suboptimally-corrected channel estimate is $\hatbsm{h}_s\triangleq\hatbsm{u}e^{\jmath\theta_s}$, where $\theta_s\triangleq \angle\left(\hat{u}_{\ell}^{*}h_{\ell}\right)$. The two channel estimates $\hatbsm{h}_s$ and $\hatbsm{h}_o$ are related by $\hatbsm{h}_s=\hatbsm{h}_oe^{\jmath(\theta_s-\theta_o)}$, and the estimation error is given by
\begin{equation}
\begin{split}
\delta\bsm{h}_s&\triangleq\hatbsm{h}_s-\bsm{h}=\hatbsm{h}_oe^{\jmath(\theta_s-\theta_o)}-\bsm{h}\\
&=\bsm{q}_oe^{\jmath(\theta_s-\theta_o)}+(e^{\jmath(\theta_s-\theta_o)}-1)\bsm{h}.
\end{split}
\end{equation}
The resulting MSE is now
\begin{equation}
\label{mse_suboptimal1}
\begin{split}
\Exp{\norm{\delta\bsm{h}_s}^2}&=\Exp{\norm{\bsm{q}_o}^2} +\Exp{|e^{\jmath(\theta_s-\theta_o)}-1|^2}\\
&=\Exp{\norm{\bsm{q}_o}^2}+2-2\Exp{\cos\left(\theta_s-\theta_o\right)}.
\end{split}
\end{equation}
The closed-form expression for $\Exp{\norm{\bsm{q}_o}^2}$ is available in Theorem~\ref{thm_optimal}, so it remains to find a closed-form for $\Exp{\cos\left(\theta_s-\theta_o\right)}$. Since $\hat{u}_{\ell}= (h_{\ell}+q_{o,\ell})e^{-\jmath\theta_o}$, we obtain
\begin{equation}
\theta_s-\theta_o=\angle\left(\hat{u}_{\ell}^{*}h_{\ell}\right)-\angle\left(\hat{{\boldsymbol u}}^{H}{\boldsymbol h}\right)=\angle\left(q_{o,\ell}^*h_{\ell}+|h_{\ell}|^2\right),
\end{equation}
It is shown in Appendix~\ref{proof_thm_optimal} that 
\begin{equation}
\label{qo_closedform}
\bsm{q}_o\simeq\frac{1}{\phantom{.}\norm{\bsm{g}}^2}\bsm{ V}\bsm{V}^{H}\delta\bsm{R}\bsm{h},
\end{equation}
where $\bsm{V}$ is a $J\times (J-1)$ matrix whose columns are the orthonormal eigenvectors of $\bsm{R}$ corresponding to the eigenvalue $\sigma^2$, and that
\begin{equation}
 \Exp{|q_{o,\ell}|^2}\simeq\frac{1}{N\norm{\bsm{g}}^4}(\sigma^2\norm{\bsm{g}}^2+\sigma^4)(1-|h_{\ell}|^2).
 \end{equation}
Using~\eqref{qo_closedform}, it can be shown that $\Exp{q_{o,\ell}}\simeq0$ and that $q_{o,\ell}$ is the sum of $N$ independent and identically distributed (i.i.d.) random variables. Furthermore, it is also shown in Appendix~\ref{proof_proper} that $\Exp{q_{o,\ell}^2}\simeq0$. Invoking the Central Limit Theorem, we can approximate $q_{o,\ell}$ by a (proper) complex Gaussian random variable with the same mean and variance. Under this approximation, the term $q_{o,\ell}^*h_{\ell}+|h_{\ell}|^2$ is also a proper complex Gaussian random variable with mean $|h_{\ell}|^2$ and variance $\frac{1}{\textstyle N\norm{\bsm{g}}^4}(\sigma^2\norm{\bsm{g}}^2+\sigma^4)(|h_{\ell}|^2-|h_{\ell}|^4)$, and furthermore, $\vartheta\triangleq\angle\left(q_{o,\ell}^*h_{\ell}+|h_{\ell}|^2\right)$ has the Ricean phase distribution~\cite{mcdonough1995} with the following probability density function
 \begin{equation} 
 \label{phase_distribution1}
 f_{\vartheta}(\vartheta)=\frac{1}{2\pi}e^{-\rho}\left(1+\sqrt{\pi\rho}\cos(\vartheta)e^{\rho\cos^2(\vartheta)}\left[1+\mbox{erf}\left(\rho\cos(\vartheta)\right)\right]\right),
 \end{equation}
 where 
 \begin{equation}
\rho=\frac{N\norm{\bsm{g}}^4|h_{\ell}|^2}{(\sigma^2\norm{\bsm{g}}^2+\sigma^4)(1-|h_{\ell}|^2)},
 \end{equation}
 and $\mbox{erf}(\cdot)$ is the error function~\cite{abramowitz}. Then, $\Exp{\cos(\vartheta)}$ is given by~\cite{linn2007}
\begin{equation}
\label{expected_cosine_exact}
\Exp{\cos(\vartheta)}=\sqrt{\frac{\pi\rho }{4}}e^{-\frac{\rho}{2}}\left[I_0\left(\frac{\rho}{2}\right)+I_1\left(\frac{\rho}{2}\right)\right],
\end{equation}
where $I_0(\cdot)$ and $I_1(\cdot)$ are Modified Bessel Functions of the First Kind of orders zero and one, respectively~\cite{abramowitz}. Going back to~\eqref{mse_suboptimal1}, we obtain the closed-form MSE expression
\begin{equation}
\Exp{\norm{\delta\bsm{h}_s}^2}\simeq \frac{1}{N\norm{\bsm {g}^4}}(\sigma^2\norm{\bsm{g}}^2+\sigma^4)(J-1)+2-2\sqrt{\frac{\pi\rho }{4}}e^{-\frac{\rho}{2}}\left[I_0\left(\frac{\rho}{2}\right)+I_1\left(\frac{\rho}{2}\right)\right].
\end{equation}
This expression can be simplified if we use the following approximation~\cite{linn2004}
\begin{equation}
\Exp{\cos(\vartheta)}\simeq e^{-\frac{1}{4\rho}}= e^{-\frac{1}{4N\norm{\bsm{g}}^4}\left(\sigma^2\norm{\bsm{g}}^2+\sigma^4\right)\left(\frac{1}{|h_{\ell}|^2}-1\right)},
\end{equation}
that yields
\begin{equation}
\label{appMSEsub}
\Exp{\norm{\delta\bsm{h}_s}^2}\simeq \frac{1}{N\norm{\bsm {g}^4}}(\sigma^2\norm{\bsm{g}}^2+\sigma^4)(J-1)+2-2e^{-\frac{1}{4N\norm{\bsm{g}}^4}\left(\sigma^2\norm{\bsm{g}}^2+\sigma^4\right)\left(\frac{1}{|h_{\ell}|^2}-1\right)}.
\end{equation}

\subsection{Largest-magnitude Phase Correction}

We can see from eq.~\eqref{appMSEsub} that the MSE decreases as the magnitude of the fading coefficient ${h}_{\ell}$ increases. For a fixed channel ${\bsm{g}}$, the MSE would be minimized when the fading coefficient with the largest magnitude is known. We let $L$ be the index of the channel coefficient with the largest magnitude, and denote by $\hatbsm{h}_a$ and $\delta\bsm{h}_a\triangleq\hatbsm{h}_a-\bsm{h}$ the resulting channel estimate when $h_L$ is known, and the corresponding estimation error, respectively. The resulting MSE is thus
\begin{equation}
\label{max_norm_mse}
\Exp{\norm{\delta\bsm{h}_a}^2}\simeq \frac{1}{N\norm{\bsm {g}^4}}(\sigma^2\norm{\bsm{g}}^2+\sigma^4)(J-1)+2-2e^{-\frac{1}{4N\norm{\bsm{g}}^4}\left(\sigma^2\norm{\bsm{g}}^2+\sigma^4\right)\left(\frac{1}{|h_{L}|^2}-1\right)}.
\end{equation}
We can obtain a simpler closed-form for $\Exp{\norm{\delta\bsm{h}_a}^2}$ by using the Taylor series expansion of the exponential term in~\eqref{max_norm_mse}. Since $|h_{L}|^2\geq 1/J$, we have that
\begin{equation}
\label{1overJ}
\frac{1}{4N\norm{\bsm{g}}^4}\left(\sigma^2\norm{\bsm{g}}^2+\sigma^4\right)\left(\frac{1}{|h_{L}|^2}-1\right)\leq\frac{J-1}{4}\left(\frac{\sigma^2}{N\norm{\bsm{g}}^2}+\frac{\sigma^4}{N\norm{\bsm{g}}^4}\right).
\end{equation}
The RHS of~\eqref{1overJ} is typically much smaller than 1, so we can accurately approximate $\Exp{\norm{\delta\bsm{h}_a}^2}$ by using the first two terms in the Taylor series expansion of the exponential term, resulting in the closed-form
\begin{equation}
\label{taylor_series}
\Exp{\norm{\delta\bsm{h}_a}^2}\simeq\frac{1}{N\norm{\bsm{g}}^4}(\sigma^{2}\norm{\bsm{g}}^2+\sigma^4)\left(J+\frac{1}{2|{h}_L|^{2}}-\frac{3}{2}\right).
\end{equation}
We will use eq.~\eqref{taylor_series} later on to obtain bounds on the probability that the WL estimator outperforms the conventional estimator under largest-magnitude phase and sign correction. 
\subsection{Training-based Phase Correction}

The previous three conventions for phase correction assume that certain information about the channel is perfectly known. In practice, however, no such information is available and pilot symbols have to be used to resolve the phase ambiguity. We will now investigate the resulting MSE if $K$ pilot symbols are used to estimate the optimal phase shift $\theta_o$, where the transmitted symbols are set to $+1$. Let ${\boldsymbol z}_k$, $k=1,\hdots, K,$ be the received sample vectors during the training period. Thus
\begin{equation}
\label{received_training}
{\boldsymbol z}_k={\bsm{g}}+{\boldsymbol n}_k,
\end{equation}
where ${\boldsymbol n}_k, k=1,\hdots, K,$ are the complex AWGN vectors with mean zero and covariance $\sigma^2{\boldsymbol I}$. To reduce the variance of the noise, we average the received vectors during the training period, obtaining
\begin{equation}
\label{average_training}
\bsm{z}_m={\bsm{g}}+\frac{1}{K}\sum_{k=1}^{K}{\boldsymbol n}_k.
\end{equation}
Using $\bsm{z}_m$, we estimate $\theta_o$ by $\hat{\theta}_o=\angle\left(\hatbsm{ u}^{H}\bsm{z}_m\right)$, and denote by $\varepsilon\triangleq \hat{\theta}_o-\theta_o$ the error in estimating $\theta_o$. We denote by $\hatbsm{h}_{t}\triangleq\hatbsm{u}e^{\jmath\hat{\theta}_o}=\hatbsm{h}_oe^{\jmath\varepsilon}$ and $\delta\bsm{h}_{t}\triangleq \hatbsm{h}_{t}-\bsm{h}=\bsm{q}_oe^{\jmath\varepsilon}+(e^{\jmath\varepsilon}-1)\bsm{h}$ the phase-corrected channel estimate, and the corresponding estimation error respectively. The resulting MSE is given by
\begin{equation}
\label{training_MSE1}
\begin{split}
\Exp{\norm{\delta\bsm{h}_{t}}^2}=\Exp{\norm{\bsm{q}_o}^2}+2-2\Exp{\cos(\varepsilon)}.
\end{split}
\end{equation}
Moreover, since $\hatbsm{u}=e^{-\jmath\theta_o}\hatbsm{h}_o$, we obtain
\begin{equation}
\label{phase_distribution0}
\begin{split}
\hatbsm{ u}^{H}\bsm{z}_m&=e^{\jmath\theta_o}\hatbsm{h}_o^{H}\left({\bsm{g}}+\frac{1}{K}\sum_{k=1}^{K}{\boldsymbol n}_k\right)= e^{\jmath\theta_o}\left(\|{\bsm{g}}\|+\frac{1}{K}\sum_{k=1}^{K}\hatbsm{h}_o^{H}{\boldsymbol{n}_k}\right),
\end{split}
\end{equation}
which means that $\varepsilon=\angle{\left(\|{\bsm{g}}\|+\frac{1}{K}\sum_{k=1}^{K}\hatbsm{ h}_o^{H}\bsm{n}_k\right)}$. The quantity $\|{\bsm{g}}\|+\frac{1}{K}\sum_{k=1}^{K}\hatbsm{ h}_o^{H}\bsm{n}_k$ is a (proper) complex Gaussian random variable with a mean of $\|{\bsm{g}}\|$ and a variance of $\sigma^2/K$, and the error $\varepsilon$ has the Ricean phase distribution~\cite{mcdonough1995} with parameter $\beta=\sqrt{\frac{K\|{\bsm{g}}\|^2}{\sigma^2}}$. Hence,
\begin{equation}
\label{expected_cosine_exact}
\Exp{\cos(\varepsilon)}=\sqrt{\frac{\pi\beta }{4}}e^{-\frac{\beta}{2}}\left[I_0\left(\frac{\beta}{2}\right)+I_1\left(\frac{\beta}{2}\right)\right]\simeq e^{-\frac{1}{4\beta}}.
\end{equation}
and
\begin{equation}
\label{total_MSE_cosine}
\begin{split}
\Exp{\norm{\delta\bsm{h}_t}^2}&=\frac{1}{N\norm{\bsm{g}}^4}(\sigma^2\norm{\bsm{g}}^2+\sigma^4)(J-1)+2\\
&-2\sqrt{\frac{\pi K\norm{\bsm{g}}^2}{4\sigma^2}}e^{-\frac{K\norm{\bsm{g}}^2}{2\sigma^2}}\left[I_0\left(\frac{K\norm{\bsm{g}}^2}{2\sigma^2}\right)+I_1\left(\frac{K\norm{\bsm{g}}^2}{2\sigma^2}\right)\right]\\
&\simeq \frac{1}{N\norm{\bsm{g}}^4}(\sigma^2\norm{\bsm{g}}^2+\sigma^4)(J-1)+2-2e^{-\frac{\sigma^2}{4K\norm{\bsm{g}}^2}}.
\end{split}
\end{equation}
As expected, the MSE under training-based phase correction approaches the one under optimal correction as the number of pilots $K$ increases.
\section{MSE Performance of the Widely Linear Estimator}
\label{MSE_performance_WL}

In this section, we will analyze the MSE performance of the WL subspace-based estimator, where the phase ambiguity is replaced by a sign ambiguity. In resolving the sign ambiguity, we will consider the same four scenarios that were considered in resolving the phase ambiguity of the conventional subspace-based estimator. Providing the same information to the WL estimator as was made available to the conventional estimator guarantees a fair comparison of the performances of the two estimators. 

\subsection{Optimal Sign Correction}

Let $b_o$ be the optimal sign correction that minimizes the Euclidean distance between the channel estimate and the true channel. It is straightforward to check that
\begin{equation}
b_o=\arg\!\min_{b\in\{\pm1\}}\norm{b\hbarbsm{u}-\barbsm{h}}^2=\mbox{sgn}\left(\hbarbsm{u}^{T}\barbsm{h}\right).
\end{equation}
We also note that $b_o=\mbox{sgn}\left\{\cos\theta_o\right\}$, where $\theta_o$ is the phase shift we used to do optimal phase correction. Denoting by $\hbarbsm{h}_o\triangleq b_o\hbarbsm{u}=[\bar{h}_{o,1},\hdots\bar{h}
_{o,2J}]$ and $\delta\barbsm{h}_o\triangleq\hbarbsm{h}_o-\barbsm{h}$ the optimally sign-corrected channel estimate and the resulting estimation error under optimal sign correction, respectively, the error $\delta\barbsm{h}_o$ can be decomposed as
\begin{equation}
\delta\barbsm{h}_o=\barbsm{q}_o+\mu_o\barbsm{h},
\end{equation}
where $\barbsm{q}_o=[\bar{q}_{o,1},\hdots,\bar{q}_{o,2J}]$ is orthogonal to $\barbsm{h}$ and $\mu_o$ is a real scalar. Hence, $\norm{\delta\barbsm{h}_o} ^2=\norm{\barbsm{q}_o}^2+|\mu_o|^2$. The following theorem presents a closed-form expression for the mean-squared error $\Exp{\norm{\delta\barbsm{h}_o} ^2}$. 
\begin{thm}
\label{thm_WL_optimal}
Under optimal sign correction, $|\mu_o|^2\ll\norm{\barbsm{q}_o}^2$, and 
\begin{equation*}
\label{optimal_WL_MSE}
\Exp{\norm{\delta\barbsm{h}_o} ^2}\simeq\Exp{\norm{\barbsm{q}_o}^2}\simeq\frac{1}{N\norm{\bsm{g}}^4}\left(\frac{\sigma^2}{2}\norm{\bsm{g}}^2+\frac{\sigma^4}{4}\right)(2J-1).
\end{equation*}
\end{thm}
The proof of the above theorem is found in Appendix~\ref{proof_thm_WL}. Since the term $\mu_o\barbsm{h}$ has a negligible impact on the estimation error $\delta\barbsm{h}_o$, we will assume for the remainder of our work that $\mu_o=0$. We will compare analytically $\Exp{\norm{\delta\barbsm{h}_o} ^2}$ and $\Exp{\norm{\delta\bsm {h}_o}^2}$ in Section~\ref{relative_performance} by evaluating the probability that $\Exp{\norm{\delta\bsm {h}_o}^2}$ is greater than $\Exp{\norm{\delta\barbsm{h}_o} ^2}$ when the statistics of the channel $\bsm{g}$ are taken into consideration. 

\subsection{Suboptimal Sign Correction}
The sign ambiguity can be resolved suboptimally when only one of the complex channel coefficients, $h_{\ell}$, is known. The complex channel coefficient $h_{\ell}$ can be expressed in terms of the real coefficients $\bar{h}_{\ell}$ and $\bar{h}_{J+\ell}$ as $h_{\ell}=\bar{h}_{\ell}+\jmath\bar{h}_{J+\ell}$. The suboptimal sign correction is given by
\begin{equation}
b_s=\mbox{sgn}\{\bar{h}_{\ell}\hat{\bar{u}}_{\ell}+\bar{h}_{J+\ell}\hat{\bar{u}}_{J+\ell}\}=\mbox{sgn}\{\cos\theta_s\},
\end{equation}
where $\theta_s$ was used to perform suboptimal phase correction in Section~\ref{section_mse_conventional}.
We denote by $\hbarbsm{h}_s\triangleq b_s\hbarbsm{u}$ and $\delta\barbsm{h}_s\triangleq\hbarbsm{h}_s-\barbsm{h}$ the suboptimally sign-corrected channel estimate, and the resulting error in the estimation of $\barbsm{h}$, respectively. To find the resulting MSE, $\Exp{\norm{\delta\barbsm{h}_s}^2}$, we first write $\hbarbsm{h}_s$ in terms of $\hbarbsm{h}_o$ as $\hbarbsm{h}_s=b_sb_o\hbarbsm{h}_o$. Hence,
\begin{equation}
\label{MSE_WL_suboptimal}
\begin{split}
\Exp{\norm{\delta\barbsm{h}_s}^2}&=\Exp{\norm{b_sb_o\hbarbsm{h}_o-\barbsm{h}}^2}\\
&=\Exp{\norm{\barbsm{q}_o}^2}+\Exp{(b_sb_o-1)^2}\\
&=\Exp{\norm{\barbsm{q}_o}^2}+4-4P\left\{b_s=b_o\right\}.
\end{split}
\end{equation}
Using the fact that $\hat{\bar{h}}_{o,\ell}=q_{o,\ell}+\bar{h}_{\ell}$, and $\hat{\bar{h}}_{o,J+\ell}=\bar{q}_{o,J+\ell}+\bar{h}_{J+\ell}$, we obtain
\begin{equation}
\begin{split}
P\left\{b_s=b_o\right\}&= P\left\{\bar{q}_{o,\ell}\bar{h}_{\ell}+\bar{h}_{\ell}^2+\bar{q}_{o,J+\ell}\bar{h}_{J+\ell}+\bar{h}_{J+\ell}^2>0\right\}\\
&=P\left\{\tilde{q}_{\ell}+|h_{\ell}|^2>0\right\},
\end{split}
\end{equation}
where $\tilde{q}_{\ell}\triangleq\bar{q}_{o,\ell}\bar{h}_{\ell}+\bar{q}_{o,J+\ell}\bar{h}_{J+\ell}$. The statistics of the random variable $\tilde{q}_{\ell}$ are needed to find a closed-form expression for $P\left\{b_s=b_o\right\}$. It is shown in Appendix~\ref{proof_thm_WL} that 
\begin{equation}
\label{qobar_closeform}
\barbsm{q}_o\simeq\frac{1}{\norm{\bsm{g}}^2}\bsm{V}_{r}\bsm{V}_{r}^{T}\delta\barbsm{R}\barbsm{h},
\end{equation}
where $\bsm{V}_{r}$ is a $2J\times(2J-1)$ matrix whose columns are the eigenvectors of $\barbsm{R}$ corresponding to the eigenvalue $\frac{\sigma^2}{2}$, and that 
\begin{equation}
\label{WLcovariance}
\Exp{\barbsm{q}_o\barbsm{q}_o^{T}}\simeq\frac{1}{N\norm{\bsm{g}}^4}\left(\frac{\sigma^2}{2}\norm{\bsm{g}}^2+\frac{\sigma^4}{4}\right)\bsm{V}_{r}\bsm{V}_{r}^{T}.
\end{equation}
It can be shown~\eqref{qobar_closeform} that $\Exp{\tilde{q}_{\ell}}\simeq0$ and that $\bar{q}_{o,\ell}$ is the sum of $N$ i.i.d. random variables. Using~\eqref{WLcovariance}, it is straightforward to show that
\begin{equation}
\label{qvariance}
\Exp{\tilde{q}_{\ell}^2}\simeq\frac{1}{N\norm{\bsm{g}}^4}\left(\frac{\sigma^2}{2}\norm{\bsm{g}}^2+\frac{\sigma^4}{4}\right)(|h_{\ell}|^2-|h_{\ell}|^4).
\end{equation}
Hence, we may invoke the Central Limit Theorem to approximate $\tilde{q}_{\ell}$ by a Gaussian random variable with the same mean and variance, thus obtaining
\begin{equation}
\label{approximate_WL_suboptimal}
P\left\{b_s=b_o\right\}\simeq Q\left(-\sqrt{\frac{N\norm{\bsm{g}}^4|h_{\ell}|^2}{\left(\frac{\sigma^2}{2}\norm{\bsm{g}}^2+\frac{\sigma^4}{4}\right)(1-|h_{\ell}|^2)}}\right),
\end{equation}
where $Q(\cdot)$ is the Guassian $Q$-function~\cite{abramowitz}. Using~\eqref{approximate_WL_suboptimal}, we finally get the MSE expression
\begin{equation}
\label{MSE_WL_suboptimal2}
\Exp{\norm{\delta\barbsm{h}_s}^2}\simeq\frac{1}{N\norm{\bsm{g}}^4}\left(\frac{\sigma^2}{2}\norm{\bsm{g}}^2+\frac{\sigma^4}{4}\right)(2J-1)+4
-4Q\left(-\sqrt{\frac{N\norm{\bsm{g}}^4|h_{\ell}|^2}{\left(\frac{\sigma^2}{2}\norm{\bsm{g}}^2+\frac{\sigma^4}{4}\right)(1-|h_{\ell}|^2)}}\right).
\end{equation}

\subsection{Largest-magnitude Sign Correction}

It is clear that the MSE expression in~\eqref{MSE_WL_suboptimal2} decreases as $|h_{\ell}|$ increases. Thus, for a fixed channel $\bsm{g}$, the MSE is lowest when the channel coefficient with the largest magnitude, $h_L$, is employed in sign correction. Denoting by $\hbarbsm{h}_a$ and $\delta\barbsm{h}_a\triangleq \hbarbsm{h}_a-\barbsm{h}$ the sign-corrected channel estimate and the resulting estimation error under largest-magnitude sign correction, respectively, the resulting MSE is given by
\begin{equation}
\label{Lmag_MSE_WL}
\begin{split}
\Exp{\norm{\delta\barbsm{h}_a}^2}&\simeq\frac{1}{N\norm{\bsm{g}}^4}\left(\frac{\sigma^2}{2}\norm{\bsm{g}}^2+\frac{\sigma^4}{4}\right)(2J-1)+4
-4Q\left(-\sqrt{\frac{N\norm{\bsm{g}}^4|h_L|^2}{\left(\frac{\sigma^2}{2}\norm{\bsm{g}}^2+\frac{\sigma^4}{4}\right)(1-|h_L|^2)}}\right).
\end{split}
\end{equation}
In this case the probability of sign recovery error becomes very small, and the above MSE expression is well approximated by the MSE under optimal sign correction in Theorem~\ref{optimal_WL_MSE}. 

\subsection{Training-based Sign Correction}

In practice, the information needed to resolve the sign ambiguity has to be estimated using training pilots. We will derive the probability of making a sign error when we use $K$ pilots to estimate the optimal sign correction $b_o$ and obtain a closed-form expression for the resulting MSE. We will also provide an expression for the unconditional probability of making a sign recovery error taking the channel statistics into account. All $K$ training symbols are set to $+1$. 

The real representation of the $K$ received sample vectors during the training period has the form
\begin{equation}
\label{received_training_sign}
\barbsm{ z}_k=\bar{\bsm{g}}+\barbsm{ n}_k,
\end{equation}
where $k=1,\hdots,K$, and $\bar{\bsm{g}}\triangleq[\Re\{{\bsm{g}}\}^{T},\Im\{{\bsm{g}}\}^{T}]^{T}$. We average the received vectors to obtain
\begin{equation}
\label{training_average_sign}
\barbsm{ z}_m=\bar{\bsm{g}}+\frac{1}{K}\sum_{k=1}^{K}\barbsm{ n}_k.
\end{equation}
Using $\barbsm{z}_m$, we estimate $b_o$ by $\hat{b}_o=\mbox{sgn}\{\hbarbsm{u}^{T}\barbsm{z}_m\}=\mbox{sgn}\left\{b_o\hbarbsm{ h}_o^{T}\barbsm{z}_m\right\}$. Hence,
\begin{equation}
\label{WL_training1}
\begin{split}
\hat{b}_o&=b_o\mbox{sgn}\left\{\hbarbsm{h}_o^{T}\barbsm{g}+\frac{1}{K}\sum_{k=1}^{K}b\hbarbsm{h}_o^{T}\barbsm{ n}_k\right\}\\
&=b_o\mbox{sgn}\left\{\norm{\bsm{g}}+\frac{1}{K}\sum_{k=1}^{K}\hbarbsm{h}_o^{T}\barbsm{ n}_k\right\}=b_o\mbox{sgn}\left\{\cos\varepsilon\right\},
\end{split}
\end{equation}
where $\varepsilon$ is the error in estimating $\theta_o$ under training-based phase correction. Denoting by $\hbarbsm{h}_t\triangleq\hat{b}_o\hbarbsm{u}$ and $\delta\barbsm{h}_t\triangleq\hbarbsm{h}_t-\barbsm{h}$ the channel estimate after sign correction and the corresponding estimation error, respectively, the resulting MSE is given by
\begin{equation}
\label{MSE_WL_training}
\Exp{\norm{\delta\barbsm{h}_t}^2}=\Exp{\norm{\barbsm{q}_o}^2}+4-4P\left\{\hat{b}_o=b_o\right\}.
\end{equation}
Moreover, we see from~\eqref{WL_training1} that 
\begin{equation}
\label{training_sign_prob}
P\left\{\hat{b}_o=b_o\right\}=P\left\{\norm{\bsm{g}}+\frac{1}{K}\sum_{k=1}^{K}\hbarbsm{h}_o^{T}\barbsm{ n}_k>0\right\}.
\end{equation}
Furthermore, the quantity $\norm{\bsm{g}}+\frac{1}{K}\sum_{k=1}^{K}\hbarbsm{h}_o^{T}\barbsm{ n}_k$ is a real Gaussian random variable with a mean of $\norm{\bsm{g}}$ and a variance of $\frac{\sigma^2}{2K}$, which means that
\begin{equation}
\label{probability_sign_error}
P\left\{\hat{b}_o\neq b_o\right\}=Q\left(\sqrt{\frac{2K\|{\bsm{g}}\|^2}{\sigma^2}}\right).
\end{equation}
Therefore, the MSE in this case is
\begin{equation} 
\label{MSE_WL_training2}
\Exp{\norm{\delta\barbsm{h}_t}^2}\simeq\frac{1}{N\norm{\bsm{g}}^4}\left(\frac{\sigma^2}{2}\norm{\bsm{g}}^2+\frac{\sigma^4}{4}\right)(2J-1)+4Q\left(\sqrt{\frac{2K\|{\bsm{g}}\|^2}{\sigma^2}}\right).
\end{equation}

Finally, since the channel coefficients $g_1,\hdots g_J$ are i.i.d. $\mathcal{CN}(0,\gamma^2)$, the unconditional probability of sign recovery which takes into account the statistics of the channel $\bsm{g}$ can be found using the Moment Generating Function (MGF) of the Gamma distribution~\cite{alouini05} and is given by
\begin{equation}
\label{average_probability_error}
P_u\left\{\hat{b}_o\neq b_o\right\}=\frac{1}{2}\left[1-\sqrt{\frac{K\gamma^2}{K\gamma^2+\sigma^2}}\sum\limits_{l=0}^{J-1}{2l\choose l}\left(\frac{\sigma^2}{4(K\gamma^2+\sigma^2)}\right)\right],
\end{equation}
for $K\gamma^2/\sigma^2>1$.

\section{Performance Comparison}
\label{relative_performance}

In this section, we compare theoretically the MSE performance of the two estimators. Our approach is to derive closed-forms and/or lower bounds on the unconditional probability that the MSE of the conventional estimator is greater than that of the WL estimator, as a function of channel length, SNR and sample size, taking into account the statistics of the channel $\bsm{g}$. We will derive a closed-form expression for this probability for the case of optimal  phase and sign correction, and two lower bounds on this probability for the case of largest-magnitude phase and sign correction. For the other two cases, suboptimal and training-based correction, such probability expressions are very difficult to derive. 

\subsection{Relative MSE Performance for Optimal Correction}
Under optimal phase and sign correction, the MSE expressions for the conventional estimator and the WL estimator are given in Theorem~\ref{thm_optimal} and Theorem~\ref{thm_WL_optimal}, respectively, and the difference in MSE is
\begin{equation}
\label{delta_MSE_optimal}
\begin{split}
\Delta\mbox{MSE}_o&\triangleq\Exp{\norm{\delta\bsm {h}_o}^2}-\Exp{\norm{\delta\barbsm{h}_o} ^2}\\
&=\frac{1}{N\norm{\bsm{g}}^4}\left(-\frac{\sigma^2}{2}\norm{\bsm{g}}^2+\sigma^4\left(\frac{J}{2}-\frac{3}{4}\right)\right).
\end{split}
\end{equation}
We are interested in evaluating $P\{\Delta\mbox{MSE}_o>0\}$, i.e., the probability that the WL estimator outperforms the conventional one. For $J\geq2$, $P\left\{\Delta\mbox{MSE}_o>0\right\}=P\left\{\norm{\bsm{g}}^2<\sigma^2\left(J-\frac{3}{2}\right)\right\}$. 
The norm-squared of the channel, $\norm{\bsm{g}}^2$, is a Central Chi-squared random variable of order $2J$, which means that~\cite{gaussian_distributions}
\begin{equation}
\label{eq_optimal_prob}
P\left\{\Delta\mbox{MSE}_o>0\right\}=G\left(\frac{\sigma^2}{\gamma^2}\left(J-\frac{3}{2}\right),J\right),
\end{equation}
where $G(x,a)$ is the regularized version of the Lower Incomplete Gamma Function defined in~\cite{abramowitz} as $G(x,s)=\frac{1}{\Gamma(s)}\int_{0}^{x}t^{s-1}e^{-t}dt$, where $\Gamma(\cdot)$ is the ordinary Gamma function. 
For fixed $s$, $G(x,s)$ is strictly increasing with respect to $x$. We will consider the case where $\frac{\gamma^2}{\sigma^2}\geq1$, which means that the average received SNR at each antenna is greater than or equal to 0 dB. In this case,
\begin{equation}
\label{gamma_inequality}
G\left(\frac{\sigma^2}{\gamma^2}\left(J-\frac{3}{2}\right),J\right)\leq G\left(J-\frac{3}{2},J\right). 
\end{equation}
Moreover, it can be shown that $G\left(J-\frac{3}{2},J\right)<\frac{1}{2}$, which means that the conventional estimator outperforms the WL estimator with probability greater than $\frac{1}{2}$. At first glance, this is rather surprising because previous works on WL subspace-based channel estimation~\cite{multicarrier05,zarifi06} report superior MSE performance for the WL estimator. However, none of the previous works considers optimal phase and sign correction. 
While the WL estimator is indeed superior in the remaining three cases (suboptimal, largest-magnitude, and training-based correction), as we will show later in this section and in Section~\ref{simulation_results}, the above result means that the superiority of the WL estimator is highly dependent on the imperfect nature of phase and sign correction. The superior performance of the WL estimator under imperfect correction makes sense because, intuitively, it is easier to correct the sign which is a binary random variable than to correct the phase which is a continuous random variable. On the other hand, when correction of both the phase and sign is optimal, the WL estimator is expected to incur more error because the dimension of its observation vector is twice that of the conventional estimator. To the best of our knowledge, we are the first to establish the relationship between the relative performance of conventional and WL subspace-based estimators and the imperfect nature of phase and sign correction.

\subsection{Relative MSE Performance for Largest-magnitude Correction}
\label{relative_MSE_largest}

Under largest-magnitude phase and sign correction, the MSE expressions for the conventional estimator and the WL estimator are given in~\eqref{taylor_series} and~\eqref{Lmag_MSE_WL}, respectively, and the difference in MSE is
\begin{equation}
\label{MSE_diff_largest}
\begin{split}
\Delta\mbox{MSE}_a&\triangleq\Exp{\norm{\delta\bsm {h}_a}^2}-\Exp{\norm{\delta\barbsm{h}_a}^2}\\
&\simeq\frac{\sigma^2}{N\norm{\bsm{g}}^2}\left(\frac{1}{2|{h}_{L}|^{2}}-1\right)+\frac{\sigma^{4}}{N\norm{\bsm{g}}^4}\left(\frac{J}{2}+\frac{1}{2|{h}_{L}|^{2}}-\frac{5}{4}\right)\\
&\phantom{==}+4Q\left(-\sqrt{\frac{N\norm{\bsm{g}}^4|h_L|^2}{\left(\frac{\sigma^2}{2}\norm{\bsm{g}}^2+\frac{\sigma^4}{4}\right)(1-|h_L|^2)}}\right)-4.
\end{split}
\end{equation}
It is very difficult to obtain a closed-form expression for $P\{\Delta\mbox{MSE}_a>0\}$ using the expression in~\eqref{MSE_diff_largest}. However, as we mentioned earlier, the probability of sign recovery error under largest-magnitude sign correction is very small and, as a result, $\Exp{\norm{\delta\barbsm{h}_a}^2}\simeq\Exp{\norm{\delta\barbsm{h}_o}^2}$, which implies that
\begin{equation}
\label{MSE_diff_largest2}
\Delta\mbox{MSE}_a\simeq\frac{\sigma^2}{N\norm{\bsm{g}}^2}\left(\frac{1}{2|{h}_L|^{2}}-1\right)+\frac{\sigma^{4}}{N\norm{\bsm{g}}^4}\left(\frac{J}{2}+\frac{1}{2|{h}_L|^{2}}-\frac{5}{4}\right)\\
\end{equation}
Using the above approximation, we obtain bounds on $P\left\{\Delta\mbox{MSE}_a>0\right\}$ that are presented in the following theorem. 
\begin{thm}
\label{approximately_optimal_thm}
The probability that the WL estimator outperforms the conventional estimator under largest-magnitude phase and sign correction can be bounded as follows:\\
Case I: $\left(J=2\right)$
\begin{equation*}
\label{upper_lower_max}
1-e^{-\frac{\sigma^2}{2\gamma^2}}<P\left\{\Delta\mbox{MSE}_a>0\right\}<1-e^{-\frac{\sigma^2}{\gamma^2}}.
\end{equation*}
Case II: $\left(J\geq3\right)$
\begin{equation*}
\label{suff_con_max3}
P\left\{\Delta\mbox{MSE}_a>0\right\}>1-J\left(\frac{1}{2}\right)^{J-1}e^{-\frac{\sigma^2}{\gamma^2}}.
\end{equation*}
\end{thm}
The proof of Theorem~\ref{approximately_optimal_thm} is found in Appendix~\ref{appendix_proof_approx_opt}.
We can see from the upper and lower bounds on $P\left\{\Delta\mbox{MSE}_a>0\right\}$ for $J=2$ that the WL estimator is better at low SNR, while the conventional estimator is better at high SNR. 

For $J\geq3$, we can use the fact that $e^{-\frac{\sigma^2}{\gamma^2}}<1$ to obtain the following lower bound which is looser but does not depend on SNR:
\begin{equation}
\label{looser_lower_bound}
P\left\{\Delta\mbox{MSE}_a>0\right\}>1-J\left(\frac{1}{2}\right)^{J-1}.
\end{equation}
Both lower bounds that apply for $J\geq3$  approach 1 as $J$ increases for fixed SNR, and the tighter one approaches 1 as SNR decreases for fixed $J$. We see from the bound in~\eqref{looser_lower_bound} that, starting with $J=4$, $P\left\{\Delta\mbox{MSE}_a>0\right\}>\frac{1}{2}$ regardless of SNR, which means that the WL estimator performs better with higher probability for this range. For $J=3$, however, which estimator is better depends on the SNR, as will be confirmed by our simulation results in Section~\ref{simulation_results}.

\section{Simulation Results}
\label{simulation_results}

We use Monte Carlo simulations to verify the accuracy of the analytical expressions we derived in the previous sections and to compare the MSE performance of the conventional and WL subspace-based estimators under the four scenarios considered in our work.

\begin{figure}[!ht]
\centering
\includegraphics[height=3.4in, width=4.5in]{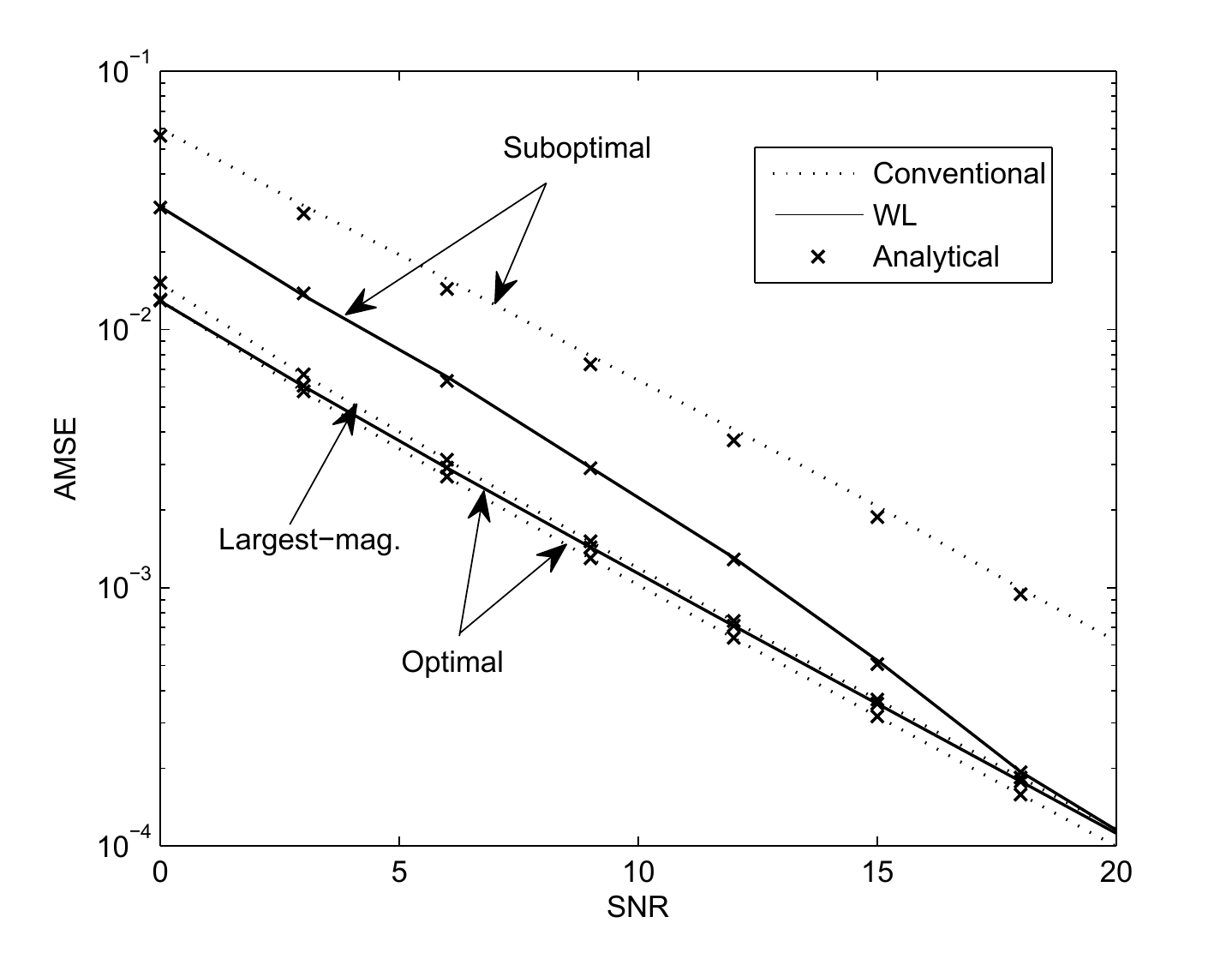}
\caption{The theoretical and experimental average MSE performance of the two estimators for the cases of optimal, suboptimal, and largest-magnitude phase and sign correction plotted vs. SNR for $J=5$ and $N=100$.}
\label{MSEvsSNR3cases}
\end{figure}

\begin{figure}[!ht]
\centering
\includegraphics[height=3.4in,width=4.5in]{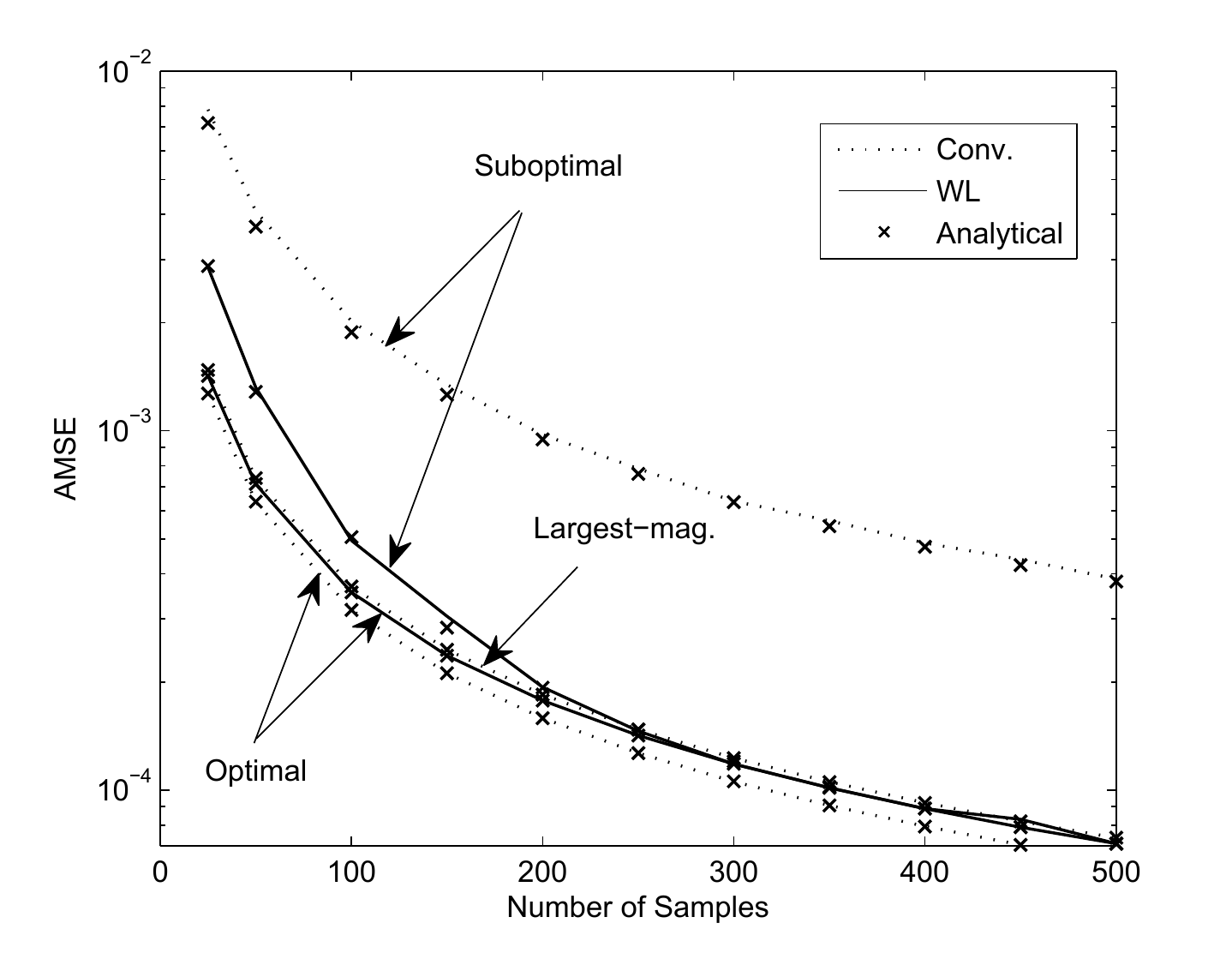}
\caption{The theoretical and experimental average MSE performance of the two estimators for the cases of optimal, suboptimal, and largest-magnitude phase and sign correction plotted vs. $N$ for $J=5$ and an SNR of 10 dB.}
\label{MSEvsN3cases}
\end{figure}

We begin with the first three cases: optimal, suboptimal, and largest-magnitude correction. Our MSE results are obtained for $J=5$ and averaged over the same set of 1000 channel realizations independently generated with $\gamma^2=1$.  We show the average MSE performance of the conventional estimator for optimal, suboptimal and largest-magnitude phase correction, and that of the WL estimator for optimal and suboptimal sign correction. The MSE performance of the WL estimator under largest-magnitude sign correction will not be shown in our plots because it is indistinguishable from the performance under optimal sign correction, due to the very low probability of making a sign recovery error in this case. In Fig.~\ref{MSEvsSNR3cases}, we plot the average MSE vs. SNR, for $N=100$, while in Fig.~\ref{MSEvsN3cases}, we plot the average MSE vs. N, for an SNR of 10 dB. Figs.~\ref{MSEvsSNR3cases} and~\ref{MSEvsN3cases} demonstrate that the derived analytical MSE expressions are highly accurate. We see that the conventional estimator slightly outperforms the WL estimator under optimal phase and sign correction. However, the WL estimator is significantly better than the conventional estimator under suboptimal correction and slightly better under largest-magnitude correction. The performance of the WL estimator under suboptimal correction approaches the performance under optimal correction for high SNR and for large sample size, which does not seem to be the case for the conventional estimator. 

\begin{figure}[!ht]
\centering
\includegraphics[height=3.4in,width=4.5in]{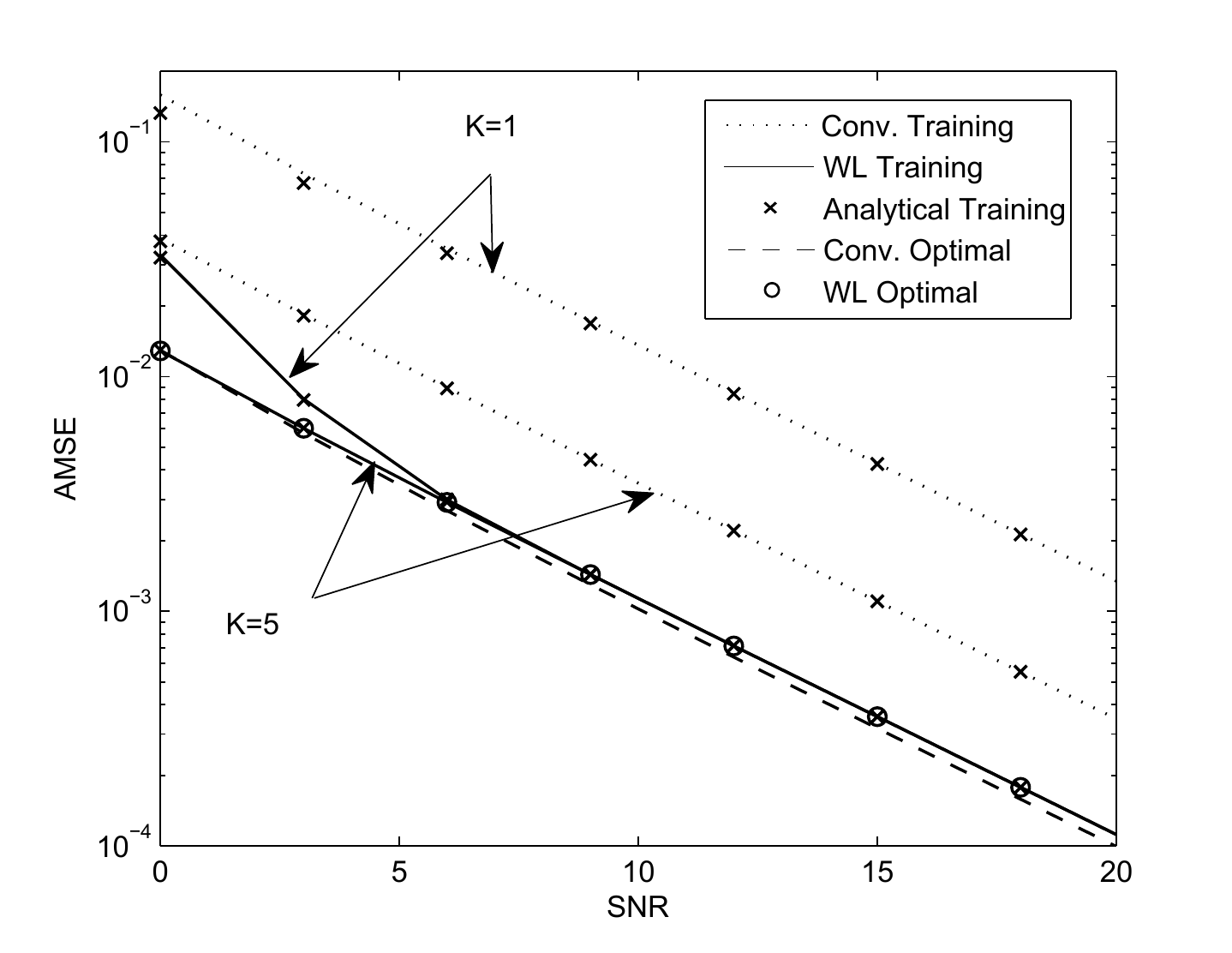}
\caption{The theoretical and experimental average MSE performance of the two estimators under training-based phase and sign correction using $K=1, 5$ for the conventional estimator and $K=1$ for the WL estimator, plotted together with optimal performance vs. SNR for $J=5$ and $N=100$.} 
\label{MSE_SNR_training}
\end{figure}

\begin{figure}[!ht]
\centering
\includegraphics[height=3.4in,width=4.5in]{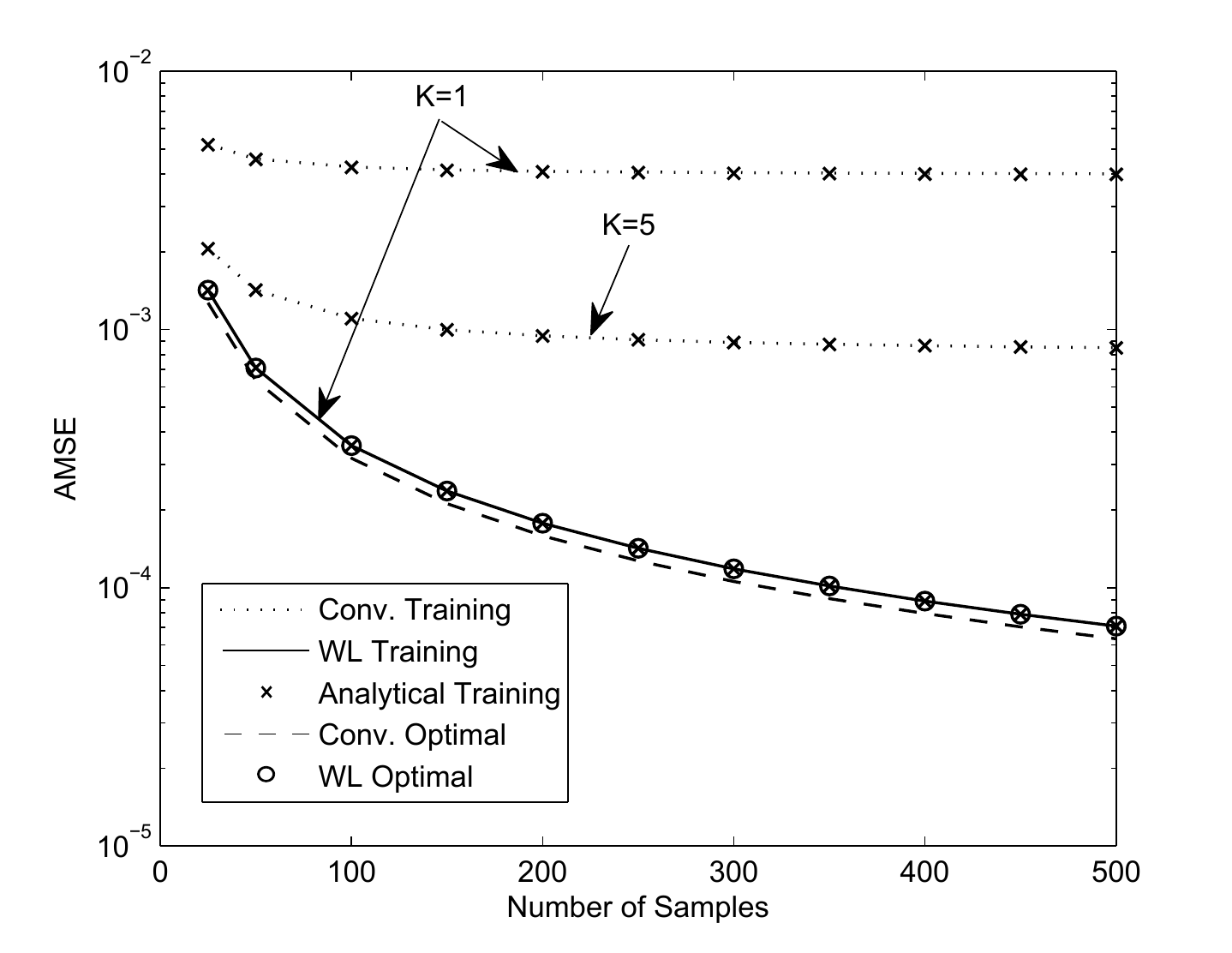}
\caption{The theoretical and experimental average MSE performance of the two estimators under training-based phase and sign correction for $K=1,5$, together with optimal performance plotted vs. $N$ for $J=5$ and an SNR of 10 dB.} 
\label{MSE_N_training}
\end{figure}

We now consider the practical case where training pilots are used for phase and sign recovery. Using the same simulation settings as before, we compare the average MSE performance of the conventional and WL estimator when pilot symbols are used to recover the phase and the sign. We study two cases, $K=1$ and $K=5$. The average MSE performance of both estimators vs. SNR under training is shown in Fig.~\ref{MSE_SNR_training}, alongside the performance under optimal phase and sign correction. As expected, the performance of both estimators improves as the number of pilots increase from 1 to 5. However, there is a huge performance gap in favor of the WL estimator. The performance of the WL estimator becomes very close to optimal performance as SNR increases for $K=1$, and becomes indistinguishable from optimal performance for $K=5$. The average MSE performance vs. $N$ is shown in Fig.~\ref{MSE_N_training} for an SNR of 10 dB. In this case, for the WL estimator we only plot the MSE for $K=1$ because it completely overlaps with the optimal performance. These results show that, in practice, the WL estimator will have superior performance because training is much more effective at recovering the sign than recovering the phase. As before, both plots demonstrate that the derived analytical expressions are highly accurate. 

\begin{figure}[!ht]
\centering
\includegraphics[height=3.4in, width=4.5in]{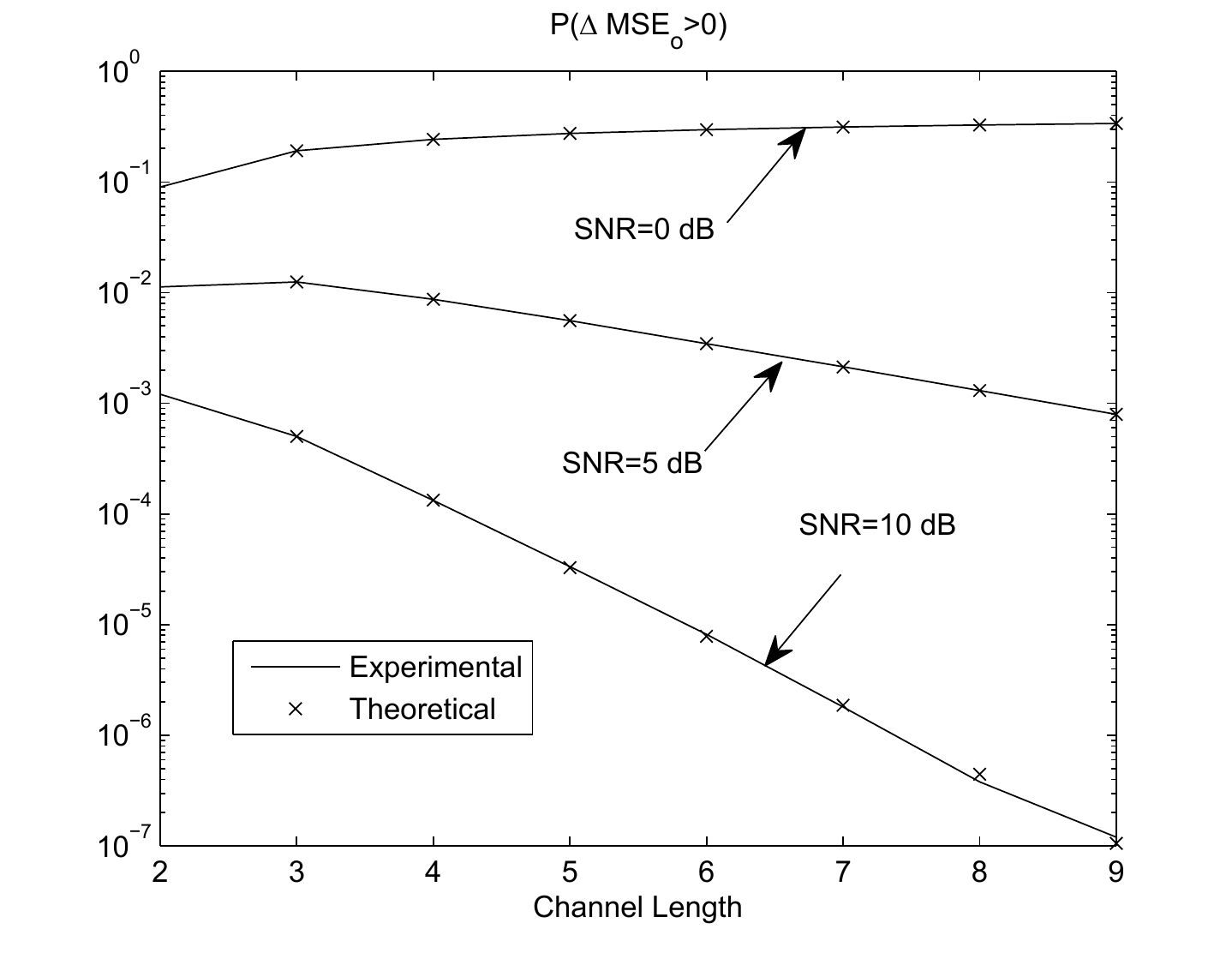}
\caption{Theoretical and experimental evaluation of the probability that the WL estimator outperforms the conventional one under optimal phase and sign correction plotted vs. $J$ for SNR values of 0, 5 and 10 dB.} 
\label{fig_prob_optimal}
\end{figure}

In Fig.~\ref{fig_prob_optimal}, we plot the analytical expression for $P\left\{\Delta\mbox{MSE}_o>0\right\}$ in~\eqref{eq_optimal_prob} and its experimental evaluation vs. channel length for SNR values of 0, 5 and 10 dB. We evaluate $P\left\{\Delta\mbox{MSE}_o>0\right\}$ experimentally using $10^7$ independently generated channel realizations (for each value of $J$) with $\gamma^2=1$. For each channel realization, we use the analytical expressions for the MSE of the two estimators to determine which estimator is superior, since we have already established the accuracy of these expressions. We see that under optimal phase and sign correction, the conventional estimator is more likely to perform better. As SNR increases, the conventional estimator performs better with overwhelmingly high probability. However, as we showed in Fig.~\ref{MSEvsSNR3cases} and Fig.~\ref{MSEvsN3cases}, the difference between the average MSE performance of the two estimators is very small. 

\begin{figure}[!ht]
\centering
\includegraphics[height=3.4in, width=4.5in]{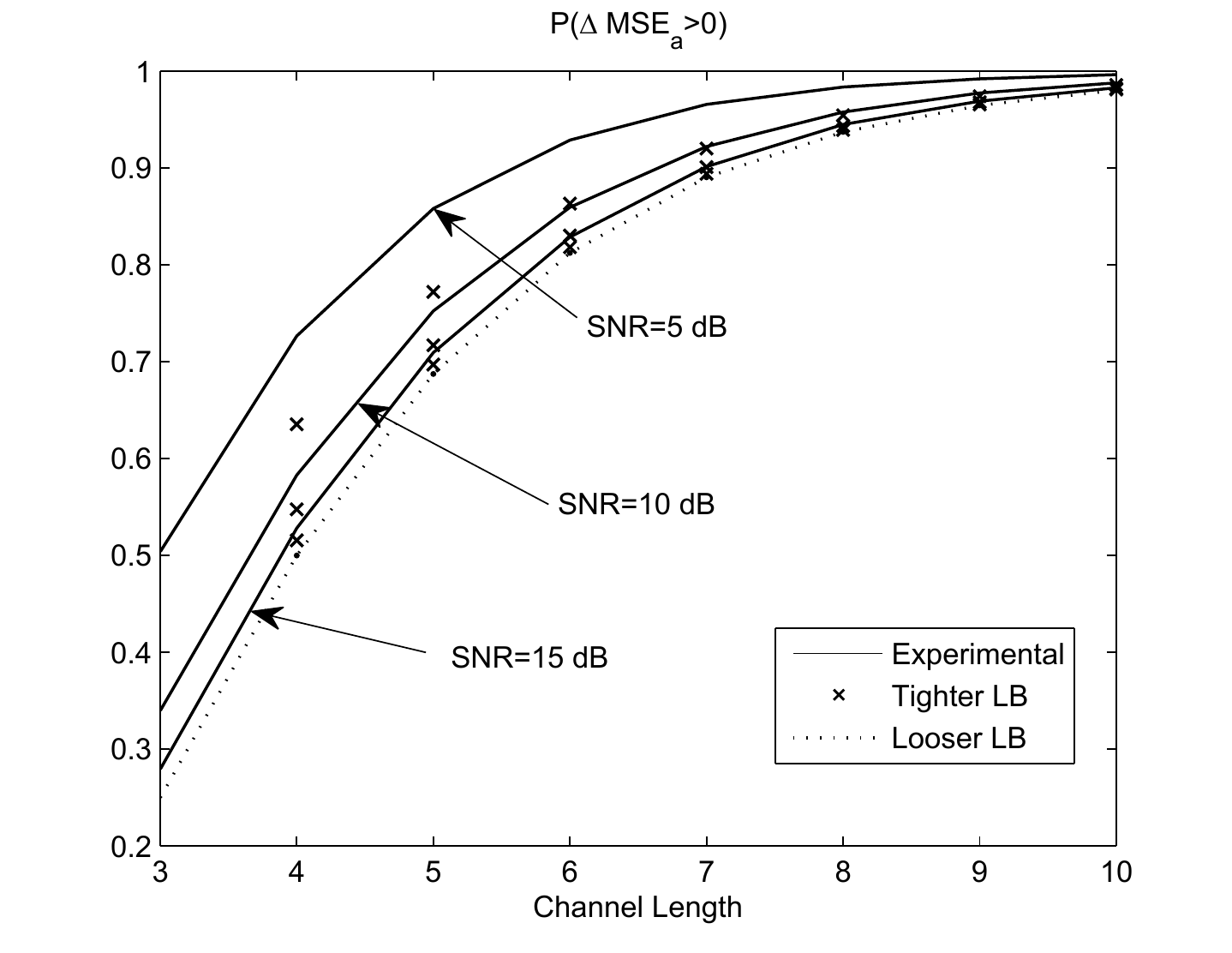}
\caption{Experimental evaluation of the probability that the WL estimator outperforms the conventional one under largest-magnitude phase and sign correction together with the lower bound in Theorem~\ref{approximately_optimal_thm} and the one in eq.~\eqref{looser_lower_bound} plotted vs. $J$ for SNR values of 5, 10 and 15 dB.} 
\label{fig_Lmag}
\end{figure}

Finally, in Fig.~\ref{fig_Lmag}, we plot the experimental evaluation of $P\left\{\Delta\mbox{MSE}_a>0\right\}$ vs. channel length for SNR values of 5, 10 and 15 dB along with the two lower bounds presented in Theorem~\ref{approximately_optimal_thm}. We evaluate $P\left\{\Delta\mbox{MSE}_a>0\right\}$ experimentally using $10^6$ independently generated channel realizations with $\gamma^2=1$. As we did for the case of optimal correction, we use the analytical expressions for the MSE to determine which estimator is superior for each channel realization. We see from Fig.~\ref{fig_Lmag} that, for $J=3$, which estimator is more likely to perform better depends on SNR, with the WL estimator being favored by low SNR and the conventional estimator being favored by high SNR. For $J\geq4$, however, the WL estimator is more likely to perform better regardless of SNR.

\section{Conclusions}
\label{conclusions}

In this paper, we presented a thorough theoretical study of the relative MSE performance of conventional
and WL subspace-based channel estimation in the context of SIMO flat-fading channels employing BPSK
modulation. Our study explicitly took into consideration the effect of covariance matrix estimation with a finite number of received samples and the impact of phase and sign ambiguity resolution. We considered four different
scenarios of phase and sign ambiguity resolution. The first three assumed that certain information
about the actual channel is perfectly known at the receiver. The first one assumed that the phase of the
inner-product between the initial channel estimate and the actual channel is known, resulting in optimal ambiguity resolution, while the second
assumed that the phase of a randomly chosen channel coefficient is known and the third one assumed that the
phase of the channel coefficient with the largest magnitude is known. We derived accurate closed-form
MSE expressions for each estimator in these cases, in addition to a closed-form expression for the
probability that the WL estimator outperforms the conventional one in the first case and a lower bound
on this probability in the third case. The three scenarios resulted in varying relative performances, with the
conventional estimator being on average slightly better in the first, and the WL estimator being significantly
better in the second and slightly better in the last. This behavior showed that the relative
performance of the two estimators is strongly related to the accuracy of phase and sign ambiguity resolution,
and that the less information about the actual channel available for ambiguity resolution, the larger the performance gap in favor
of the WL estimator. We also studied the more realistic scenario where pilot symbols are used to resolve
the two ambiguities. We derived accurate closed-form expressions for the MSE of the two estimators
assuming that the same number of pilots are used. Our simulations showed that in this when pilots are used there is
a significant performance gap in favor of the WL estimator which performs very close to optimal
even with a single pilot. This scenario showed that the WL estimator significantly outperforms the
conventional one when the information about the channel which is available for ambiguity resolution is inaccurate.

\appendices

\section{Proof of Theorem~\ref{thm_optimal}}
\label{proof_thm_optimal}
The phase-corrected channel estimate is $\hatbsm{h}_o=\hatbsm{u}e^{\jmath\angle\left(\hatbsm{u}^{H}\bsm{h}\right)}$. Hence, $\bsm{h}^{H}\hatbsm{h}_o=|\bsm{h}^{H}\hatbsm{u}|\geq0$. Since $\hatbsm{h}_o=\bsm{h}+\bsm{q}_o+\alpha_o\bsm{h}$, we have that $\bsm{h}^{H}\hatbsm{h}_o=1+\alpha_o$, which implies that $\alpha_o$ is real and $\alpha_o\geq-1$. Using the fact that $\hatbsm{h}_o^{H}\hatbsm{h}_o=1$, we obtain
\begin{equation}
\alpha_o^2+2\alpha_o+\norm{\bsm{q}_o}^2=0.
\end{equation}
Since $\alpha_o\geq-1$, the solution of the above quadratic equation is $\alpha_o=-1+\sqrt{1-\norm{\bsm{q}_o}^2}$. Moreover, since $\norm{\bsm{q}_o}^2<1$, we can use the Taylor series expansion $\sqrt{1-\norm{\bsm{q}_o}^2}=1-\frac{1}{2}\norm{\bsm{q}_o}^2+O\left(\norm{\bsm{q}_o}^4\right)$, to obtain $\alpha_o=-\frac{1}{2}\norm{\bsm{q}_o}^2+O\left(\norm{\bsm{q}_o}^4\right)$, which means that $|\alpha_o|^2=\frac{1}{4}\norm{\bsm{q}_o}^4+O\left(\norm{\bsm{q}_o}^6\right)$. Therefore, $|\alpha_o|^2\ll\norm{\bsm{q}_o}^2$, $\delta\bsm{h}_o\simeq\bsm{q}_o$ and $\Exp{\norm{\delta\bsm{h}_o}^2}\simeq\Exp{\norm{\bsm{q}_o}^2}$.

To find $\Exp{\norm{\bsm{q}_o^2}}$, we denote by $\hat{\lambda}_{1}$ the largest eigenvalue of $\hatbsm{R}$, and let $\delta\lambda_{1}\triangleq\hat{\lambda}_{1}-\lambda_{1}$. Since $\hatbsm{R}\hatbsm{h}_o=\hat{\lambda}_{1}\hatbsm{h}_o$, we have
\begin{equation}
\begin{split}
\label{eq_start_linear}
(\bsm{R}+\delta\bsm{R})(\bsm{h}+ \bsm{q}_o)&=(\lambda_{1}+\delta\lambda_{1})(\bsm{h}+ \bsm{q}_o).
\end{split}
\end{equation}  
Using the fact that $\bsm{ R}\bsm{h}=\lambda_{1}\bsm{h}$, and ignoring the second-order perturbation terms $\delta\bsm{R}\bsm{q}_o$ and $\delta\lambda_{1}\bsm{q}_o$ in~\eqref{eq_start_linear} (as in~\cite{friedlander98}), we obtain
\begin{equation}
\label{eq01}
(\bsm{R}-\lambda_{1}\bsm{I})\bsm{q}_o\simeq-\delta\bsm{R}\bsm{h}+\delta\lambda_{1}\bsm{h}.
\end{equation}
Furthermore, we have that $(\bsm{R}-\lambda_{1}\bsm{I})=-\norm{\bsm g}^2\bsm{V}\bsm{V}^H$, where $\bsm{V}$ is a $J\times (J-1)$ matrix whose columns are the orthonormal eigenvectors of $\bsm{R}$ corresponding to the eigenvalue $\sigma^2$ of multiplicity $J-1$ (i.e., the noise subspace).  Multiplying the two sides of~\eqref{eq01} by $\bsm{V}\bsm{V}^H$, we obtain
\begin{equation}
\label{eq011}
\bsm{q}_o\simeq\frac{1}{\phantom{.}\norm{\bsm{g}}^2}\bsm{ V}\bsm{V}^{H}\delta\bsm{R}\bsm{h}.
\end{equation}
From eq.~\eqref{eq011}, we see that $\Exp{\bsm{q}_o}\simeq\bsm{0}$. Let $\bsm{Q}\triangleq\Exp{\bsm{q}_o\bsm{q}_o^{H}}$, the MSE is $\Exp{\norm{\bsm{q}_o}^2}=\mbox{tr}\left\{\bsm{Q}\right\}$, and we can expand $\bsm{Q}$ as
\begin{equation}
\label{Q}
\bsm{Q}\simeq\frac{1}{\phantom{.}\norm{\bsm{g}}^4}\Expected\{\bsm{V}\bsm{V}^{H}\delta\bsm{R}\bsm{h}\bsm{h}^{H}\delta\bsm{R}\bsm{V}\bsm{V}^{H}\}.
\end{equation}
To obtain a closed-form for the above expectation, we must evaluate the term $\Expected\{\delta\bsm{R}\bsm{h}\bsm{h}^{H}\delta\bsm{R}\}$. Using the approach followed in~\cite{xu02}, it can be shown that
\begin{equation}  
\label{mixed_gaussian}
\Expected\{\delta\bsm{R}\bsm{h}\bsm{h}^{H}\delta\bsm{R}\}=\frac{1}{N}[\sigma^2\norm{\bsm{g}}^2\phantom{.}\bsm{h}\bsm{h}^{H}+(\sigma^2\hspace{-0.025ex}\norm{\bsm{g}}^2+\sigma^4)\bsm{I}].
\end{equation}
Therefore, 
\begin{equation}
\label{expression_Q}
\bsm{Q}\simeq\frac{1}{N\norm{\bsm{g}}^4}(\hspace{0.1ex}\sigma^2\norm{\bsm{g}}^2+\sigma^4)\bsm{V}\bsm{V}^{H}.
\end{equation}
From the above equation, we see that, for any $\ell=1,\hdots, J$, $\Exp{q_{o,\ell}}=\frac{1}{N\norm{\bsm{g}}^4}(\hspace{0.1ex}\sigma^2\norm{\bsm{g}}^2+\sigma^4)(1-|h_{\ell}|^2)$.
Finally, the MSE is given by
\begin{equation}
\label{fixed_error_expectation}
\Exp{\norm{\bsm{q}_o}^2}\simeq\mbox{tr}\left\{\frac{1}{N\norm{\bsm{g}}^4}(\sigma^2\norm{\bsm{g}}^2+\sigma^4)\bsm{V}\bsm{V}^{H}\right\}=\frac{1}{N\norm{\bsm{g}}^4}(\sigma^2\norm{\bsm{g}}^2+\sigma^4)(J-1).
\end{equation}

\section{}
\label{proof_proper}
In this appendix, we show that $\Expected\{q_{o,\ell}^{2}\}\simeq0$ for $\ell=1,\hdots, J$. We first let
$\bsm{W}\triangleq\Expected\{\bsm{q}\bsm{q}^{T}\}\simeq\Expected\{\bsm{V}\bsm{V}^{H}\delta\bsm{R}\bsm{h}\bsm{h}^{T}\delta\bsm{R}^{*}\bsm{V}^{*}\bsm{V}_{l}^{T}\}$. Hence $\Expected\{q_{o,\ell}^{2}\}=\bsm{W}(\ell,\ell)$, where $\bsm{W}(i,j)$ is the $(i,j)$th entry of $\bsm{W}$. To obtain a closed-form expression for $\bsm{W}$, we need to evaluate $\Expected\{\delta\bsm{R}\bsm{h}\bsm{h}^{T}\delta\bsm{R}^{*}\}$. It is easy to see that
 \begin{equation}
\label{eqa01}
\Expected\{\delta\bsm{R}\bsm{h}\bsm{h}^{T}\delta\bsm{R}^{*}\}=\frac{1}{N}(\Expected\{\bsm{r}(i)\bsm{r}(i)^{H}\bsm{h}\bsm{h}^{T}\bsm{r}(i)^{*}\bsm{r}(i)^{T}\}-\bsm{R}\bsm{h}^{T}\bsm{R}^{*}),
\end{equation}
where $\bsm{r}(i)$ is the $i$th received vector sample. Moreover,
\begin{equation}
\label{eqa02}
\begin{split}
\Expected\{\bsm{r}(i)\bsm{r}(i)^{H}\bsm{h}\bsm{h}^{T}\bsm{r}(i)^{*}\bsm{r}(i)^{T}\}=&\norm{\bsm{g}}^4\bsm{h}\bsm{h}^{T}+4\sigma^{2}\norm{\bsm{g}}^2\bsm{h}\bsm{h}^{T}+\Expected\{\bsm{n}(i)\bsm{n}(i)^{H}\bsm{h}\bsm{h}^{T}\bsm{n}(i)^{*}\bsm{n}(i)^{T}\}.
\end{split}
\end{equation}
To find $\Expected\{\bsm{n}(i)\bsm{n}(i)^{H}\bsm{h}\bsm{h}^{T}\bsm{n}(i)^{*}\bsm{n}(i)^{T}\}$, we use the property~\cite{theory_matrices}
\begin{equation}
\mbox{vec}(\bsm{A}\bsm{X}\bsm{B})=(\bsm{B}^{T}\otimes\bsm{A})\mbox{vec}(\bsm{X}), 
\end{equation}
where the $\mbox{vec}(\cdot)$ operator stacks the column vectors of a matrix below one another and $\otimes$ denotes the Kronecker product. Hence, we obtain
\begin{equation}
\label{eqa0}
\Expected\{\bsm{n}(i)\bsm{n}(i)^{H}\bsm{h}\bsm{h}^{T}\bsm{n}(i)^{*}\bsm{n}(i)^{T}\}=\mbox{unvec}[\Expected\{\bsm{n}(i)\bsm{n}(i)^{H}\otimes\bsm{n}(i)\bsm{n}(i)^{H}\}\mbox{vec}(\bsm{h}\bsm{h}^{T})],
\end{equation}
where $\mbox{unvec}(\cdot)$ converts a vector into matrix format by ordering the elements of the vector into columns of equal length. 
It is straightforward to check that $\Expected\{\bsm{n}(i)\bsm{n}(i)^{H}\otimes\bsm{n}(i)\bsm{n}(i)^{H}\}=\sigma^{4}\bsm{I}_{J^2}+\sigma^4\bsm{A}$ where $\bsm{A}$ is a $J^2\times J^2$ matrix made of  $J\times J$ subblocks each of dimension $J\times J$, with the $(i,j)th$ sub-block given by $\bsm{s}_j\bsm{s}_{i}^{T}$ where $\bsm{s}_{i}$ is the $J\times1$ vector whose \textit{i}th entry is 1 and all other entries are zero. Thus,
\begin{equation}
\label{eqa04}
\begin{split}
\Expected\{\bsm{n}(i)\bsm{n}(i)^{H}\bsm{h}\bsm{h}^{T}\bsm{n}(i)^{*}\bsm{n}(i)^{T}\}&=\mbox{unvec}[(\sigma^{4}\bsm{I}_{J^2}+\sigma^4\bsm{A})\mbox{vec}(\bsm{h}\bsm{h}^{T})]=2\sigma^4\bsm{h}\bsm{h}^{T},
\end{split}
\end{equation}
where we have used eq. (39) from~\cite{xu02}. Combining~\eqref{eqa01},~\eqref{eqa02}, and~\eqref{eqa04}, we obtain
\begin{equation}
\label{eq11}
\begin{split}
\Expected\{\delta\bsm{R}\bsm{h}\bsm{h}^{T}\delta\bsm{R}^{*}\}&=\frac{1}{N}(\norm{\bsm{g}}^4+4\sigma^{2}\norm{\bsm{g}}^2+2\sigma^4)\bsm{h}\bsm{h}^{T}-\frac{1}{N}(\norm{\bsm{g}}^4+2\sigma^2\norm{\bsm{g}}^2+\sigma^{4})\bsm{h}\bsm{h}^{T}\\
&=\frac{1}{N}(2\sigma^{2}\norm{\bsm{g}}^2+\sigma^{4})\bsm{h}\bsm{h}^{T}.
\end{split}
\end{equation}
Therefore, $\bsm{W}\simeq\frac{1}{N}(2\sigma^{2}\norm{\bsm{g}}^2+\sigma^{4})\bsm{V}\bsm{V}^{H}\bsm{h}\bsm{h}^{T}\bsm{V}^{*}\bsm{V}^{T}=0$, since all columns of $\bsm{V}$ are orthogonal to $\bsm{h}$. Therefore, $\Expected\{q_{o,\ell}^{2}\}=\bsm{W}(\ell,\ell)\simeq0$.

\section{Proof of Theorem~\ref{thm_WL_optimal}}
\label{proof_thm_WL}
Following similar steps to those in Appendix~\ref{proof_thm_optimal}, we can show that $|\mu_0|^2\ll\norm{\barbsm{q}_o}^2$ and that
\begin{equation}
\label{eq35}
\delta\barbsm{h}_o\simeq\barbsm{q}_o\simeq\frac{1}{\norm{\bsm{g}}^2}\bsm{V}_{r}\bsm{V}_{r}^{T}\delta\barbsm{R}\barbsm{h}.
\end{equation}
where $\bsm{V}_{r}$ is a $2J\times(2J-1)$ matrix whose columns are the eigenvectors of $\barbsm{R}$ corresponding to the eigenvalue $\frac{\sigma^2}{2}$ of multiplicity $2J-1$ (i.e., the noise subspace). Thus, $\Exp{\barbsm{q}_o}\simeq0$. Let
\begin{equation}
\label{WLQ}
\bar{\bsm{Q}}\triangleq\Exp{\barbsm{ q}_o\barbsm{ q}_o^{T}}\simeq\frac{1}{\norm{\bsm{g}}^4}\{\bsm{V}_{r}\bsm{V}_{r}^{T}\Expected\{\delta\barbsm{R}\barbsm{h}\barbsm{h}^{T}\delta\barbsm{R}\}\bsm{V}_{r}\bsm{V}_{r}^{T}\}.
\end{equation}
By applying the same approach used in~\cite{xu02} for real signals, we can find the following closed form for $\Expected\{\delta\barbsm{R}\barbsm{h}\barbsm{h}^{T}\delta\barbsm{R}\}$ (the derivation has been omitted for brevity): 
\begin{equation}
\label{sm:lem_eqn_wl}
\Exp{\delta\barbsm{R}\barbsm{h}\barbsm{h}^{T}\delta\barbsm{R}}=\frac{1}{N}\left(\left(\frac{3}{2}\sigma^2\norm{\bsm{g}}^2+\frac{\sigma^4}{4}\right)\barbsm{h}\barbsm{h}^{T}+\left(\frac{\sigma^2}{2}\norm{\bsm{g}}^2+\frac{\sigma^4}{4}\right)\bsm{I}\right).
\end{equation}
Substituting~\eqref{sm:lem_eqn_wl} into~\eqref{WLQ}, we obtain
\begin{equation}
\label{WLQ2}
\begin{split}
\barbsm{Q}&\simeq\frac{1}{N\norm{\bsm{g}}^4}\left(\frac{\sigma^2}{2}\norm{\bsm{g}}^2+\frac{\sigma^4}{4}\right)\bsm{V}_{r}\bsm{V}_{r}^{T}.
\end{split}
\end{equation}
Finally, the MSE is given by
\begin{equation}
\label{eq46_appendix}
\begin{split}
\Exp{\norm{\delta\barbsm{h}_o}^2}\simeq\mbox{tr}\{\bar{\bsm{Q}}\}&\simeq\frac{1}{N\norm{\bsm{g}}^4}\left(\frac{\sigma^2}{2}\norm{\bsm{g}}^2+\frac{\sigma^4}{4}\right)(2J-1).
\end{split}
\end{equation}

\section{Proof of Theorem~\ref{approximately_optimal_thm}}
\label{appendix_proof_approx_opt}
The difference in MSE for under largest-magnitude correction is
\begin{equation}
\label{eq48}
\begin{split}
\Delta\mbox{MSE}_a&\simeq\frac{\sigma^2}{N\norm{\bsm{g}}^2}\left(\frac{1}{2|{h}_{L}|^{2}}-1\right)+\frac{\sigma^{4}}{N\norm{\bsm{g}}^4}\left(\frac{J}{2}+\frac{1}{2|{h}_L|^{2}}-\frac{5}{4}\right).
\end{split}
\end{equation}
Substituting $h_L=\frac{\textstyle{g}_L}{\textstyle\norm{\bsm{g}}}$, the WL estimator outperforms the linear one when
\begin{equation}
\label{eq49_1}
\frac{\norm{\bsm{g}}^4}{2|{g}_L|^{2}}-\norm{\bsm{g}}^2+\frac{\sigma^2\norm{\bsm{g}}^2}{2|{g}_L|^{2}}>\sigma^2\left(\frac{5}{4}-\frac{J}{2}\right).
\end{equation}
We first consider the case $J=2$, and let $p$ be the index of the channel coefficient with the smallest magnitude. The condition in~\eqref{eq49_1} simplifies to
\begin{equation}
\label{max1}
|{g}_{p}|^{4}-|{g}_L|^{4}+\sigma^2|{g}_{p}|^2+\frac{\sigma^2}{2}|{g}_L|^2>0.
\end{equation}
It is easy to see from~\eqref{max1} that 
\begin{equation}
\label{max2}
P\left(\Delta\mbox{MSE}_a>0\right)<1-P(|{g}_L|^{2}-|{g}_{p}|^{2}\geq\sigma^2).
\end{equation}
Moreover, ${g}_1$ and ${g}_2$ are i.i.d., so $P(L=1)=P(L=2)=\frac{1}{2}$ and
\begin{equation}
\label{max_2_iid}
\begin{split}
P\left(|{g}_L|^{2}-|{g}_{p}|^{2}\geq\sigma^2\right)&=\frac{1}{2}P\left(|{g}_1|^2-|{g}_2|^2\geq\sigma^2\big\vert L=1\right)+\frac{1}{2}P\left(|{g}_2|^2-|{g}_1|^2\geq\sigma^2\big\vert L=2\right)\\
&=P\left(|{g}_1|^2-|{g}_2|^2\geq\sigma^2\big\vert L=1\right)\\
&=2P\left(|{g}_1|^2-|{g}_2|^2\geq\sigma^2\right).
\end{split}
\end{equation}
Since ${g}_1$ and ${g}_2$ are i.i.d. $\mathcal{CN}(0,\gamma^2)$, we have that~\cite{gaussian_distributions} $P(|{g}_1|^{2}-|{g}_2|^{2}\geq\sigma^2)=\frac{1}{2}e^{-\frac{\sigma^2}{\gamma^2}}$, which provides us with the upper bound
\begin{equation}
P\left(\Delta\mbox{MSE}_a>0\right)<1-e^{-\frac{2\sigma^2}{\gamma^2}}.
\end{equation}
Moreover, it can also be inferred from~\eqref{max1} that
\begin{equation}
\label{second_inference}
P\left(\Delta\mbox{MSE}_a>0\right)>1-P\left(|{g}_L^{2}-|{g}_{p}|^{2}\geq\frac{\sigma^2}{2}\right),
\end{equation}
which allows us to lower-bound $P\left(\Delta\mbox{MSE}_a>0\right)$ by $1-e^{-\frac{\sigma^2}{2\gamma^2}}$. Therefore,
\begin{equation}
\label{upper_lower_max_appendix}
1-e^{-\frac{\sigma^2}{\gamma^2}}<P\left(\Delta\mbox{MSE}_a>0\right)<1-e^{-\frac{2\sigma^2}{\gamma^2}}.
\end{equation}

We will now consider the case $J\geq3$. In this case, the RHS of eq.~\eqref{eq49_1} is negative, and a sufficient condition for the WL estimator to be better is
\begin{equation}
\label{sufficient_condition}
2|{g}_L|^{2}-\norm{\bsm{g}}^2\leq\sigma^2.
\end{equation}
It is convenient to define the random variable $D^{(i)}\triangleq2|{g}_{i}|^{2}-\norm{\bsm{g}}^2$, for $i=1,\hdots,J$. The random variable $D^{(i)}$ is the difference of two independent central Chi-square random variables of degrees $2$ and $2J-2$, respectively, and $P\left(D^{(i)}\geq x\right)=\left(\frac{1}{2}\right)^{J-1}e^{\frac{-x}{\gamma^2}}$ for any $x\geq0$~\cite{gaussian_distributions}. 
Moreover, for any $i=1,\hdots, J$,
\begin{equation}
\label{rand2max}
\begin{split}
P\left(D^{(i)}\geq\sigma^2\right)&=P\left(D^{(i)}\geq\sigma^2\big\vert L=i\right)P(L=i)+P\left(D^{(i)}\geq\sigma^2\big\vert L\neq i\right)P(L\neq i)\\
&=P\left(D^{(i)}\geq\sigma^2\big\vert L=i\right)P(L=i)\\
&=\frac{1}{J}P\left(D^{(i)}\geq\sigma^2\big\vert L=i\right).
\end{split}
\end{equation}
Thus, $P\left(D^{(i)}\geq\sigma^2\big\vert L=i\right)=JP\left(D^{(i)}\geq\sigma^2\right)=J\left(\frac{1}{2}\right)^{J-1}e^{\frac{-\sigma^2}{\gamma^2}}$.
Going back to the sufficient condition in~\eqref{sufficient_condition}, we obtain
\begin{equation}
\label{suff_con_max}
\begin{split}
P\left(\Delta\mbox{MSE}_a>0\right)&> 1-P\left(2|{g}_L|^{2}-\norm{\bsm{g}}^2\geq\sigma^2\right)\\
&=1-\sum\limits_{i=1}^{J}P\left(2|{g}_i|^{2}-\norm{\bsm{g}}^2\geq\sigma^2\big\vert L=i\right)P\left(L=i\right)\\
&=1-\sum\limits_{i=1}^{J}P\left(D^{(i)}\geq\sigma^2\big\vert L=i\right)P\left(L=i\right)\\
&=1-J\left(\frac{1}{2}\right)^{J-1}e^{\frac{-\sigma^2}{\gamma^2}}.
\end{split}
\end{equation}

\bibliographystyle{IEEEtran}
\bibliography{IEEEabrv,widelylinearbib}

\newpage

\end{document}